\documentclass[preprint,3p,times,onecolumn]{elsarticle}

\journal{Combustion and Flame}

\usepackage{latexsym,amssymb,graphicx,deluxetable,setspace}

\def\eps@scaling{1.0}%
\newcommand\epsscale[1]{\gdef\eps@scaling{#1}}%
\newcommand\plotone[1]{%
 \centering 
 \leavevmode 
 \includegraphics[width={\eps@scaling\textwidth}]{#1}%
}%
\newcommand\plottwo[2]{%
 \centering 
 \leavevmode 
 \textwidth=.45\textwidth 
 \includegraphics[width={\eps@scaling\textwidth}]{#1}%
 \hfil
 \includegraphics[width={\eps@scaling\textwidth}]{#2}%
}%
\newcommand\plotfiddle[7]{%
 \centering 
 \leavevmode 
 \vbox\@to#2{\rule{\z@}{#2}}%
 \includegraphics[%
  scale=#4, 
  angle=#3, 
  origin=c 
 ]{#1}%
}%

\newif\ifAMStwofonts

\ifAMStwofonts \else 

\fi

\newcommand{\pd}[2]{ { \partial {#1} \over \partial {#2} } }

\newcommand{\vect}[1]{\mathbf{#1}}   

\newcommand{\sfrac}[2]{\,{}^{#1}\!/_{#2}}

\newcommand{\beq}{\begin{equation}}
\newcommand{\eeq}{\end{equation}}
\newcommand{\bdm}{\begin{displaymath}}
\newcommand{\edm}{\end{displaymath}}
\newcommand{\ie}{i.e., }
\newcommand{\ea}{et al.}
\newcommand{\eg}{e.g., }

\begin{document}

\begin{frontmatter}



\title{The Interaction of High-Speed Turbulence with Flames: Global
Properties and Internal Flame Structure}

\author{A.Y.~Poludnenko\corref{ayp}}
\ead{apol@lcp.nrl.navy.mil}
\author{E.S.~Oran}
\cortext[ayp]{Corresponding author. Fax: +1 202 767 4798.}
\address{Laboratory for Computational Physics and Fluid Dynamics,
Naval Research Laboratory, Washington, DC 20375, USA}

\begin{abstract}
We study the dynamics and properties of a turbulent flame, formed in
the presence of subsonic, high-speed, homogeneous, isotropic
Kolmogorov-type turbulence in an unconfined system. Direct numerical
simulations are performed with Athena-RFX, a massively parallel, fully
compressible, high-order, dimensionally unsplit, reactive-flow code. A
simplified reaction-diffusion model represents a stoichiometric
H$_2$-air mixture. The system being modeled represents turbulent
combustion with the Damk\"ohler number $Da = 0.05$ and with the
turbulent velocity at the energy injection scale 30 times larger than
the laminar flame speed. The simulations show that flame interaction
with high-speed turbulence forms a steadily propagating turbulent
flame with a flame brush width approximately twice the energy
injection scale and a speed four times the laminar flame speed. A
method for reconstructing the internal flame structure is described
and used to show that the turbulent flame consists of tightly folded
flamelets. The reaction zone structure of these is virtually identical
to that of the planar laminar flame, while the preheat zone is
broadened by approximately a factor of two. Consequently, the system
evolution represents turbulent combustion in the thin-reaction zone
regime. The turbulent cascade fails to penetrate the internal flame
structure, and thus the action of small-scale turbulence is suppressed
throughout most of the flame. Finally, our results suggest that for
stoichiometric H$_2$-air mixtures, any substantial flame broadening by
the action of turbulence cannot be expected in all subsonic regimes.

\end{abstract}

\end{frontmatter}



\section{Introduction}
\label{Intro}

In the last twenty years, studies of premixed turbulent combustion
have witnessed an explosive growth. The motivation for this comes from
a remarkably broad range of both engineering and basic science
applications -- from the design of internal combustion engines, to
problems of ensuring safety in mines, to the dynamics and properties
of turbulent thermonuclear flames powering type Ia supernovae. Such
rapid development has been enabled by significant advances in
experimental techniques, in particular in terms of the diagnostics, as
well as by the substantial increase in the capabilities of the
computational infrastructure and numerical algorithms, all of which
made direct numerical simulation (DNS) a viable, and often
indispensable, tool for turbulent combustion research.

Despite the significant increase in the body of experimental and
numerical data, the overall paradigm still prevailing today is
reflected in a variety of combustion regime diagrams (e.g., see
\cite{Peters,Williams,Williams2} and references therein), which attempt
to provide a comprehensive classification of the different modes of
turbulence-flame interaction. The key underlying assumption is that
such classification can be made based on a very limited set of
parameters, namely, the characteristic large-scale turbulent velocity
and the spatial scale. With this ansatz, the diagrams can be
constructed by comparing various turbulent timescales, \eg eddy
turnover timescales on the integral and Kolmogorov scales, with those
of the unperturbed laminar flame. The resulting classifications
include such regimes as ``wrinkled'' and ``corrugated flamelets,'' as
well as ``thin,'' ``broken,'' and ``distributed reaction zones,''
which suggest the typical qualitative structure of the turbulent flame
under specific conditions.

The basis for this paradigm is the picture which was first suggested
by Damk\"ohler almost 70 years ago \cite{Damkohler,Driscoll,Bilger}.
Large-scale motions are responsible for the overall stretching and
folding of the flame that constitutes the flame brush. This process
increases the flame surface area, which directly determines the global
properties of the turbulent flame, such as its width and speed.
Therefore, depending on whether the characteristic timescale of these
large-scale motions is greater or smaller than the laminar flame
self-crossing time, the flame is either able or not to reorganize and
adjust itself to the action of turbulence.

On the other hand, small scales penetrate inside individual flamelets,
thus affecting their internal structure and broadening their preheat
and reaction zones. The efficiency of this process is controlled by
the magnitude of the Kolmogorov scale with respect to the laminar
flame width, $\delta_L$, or the size of the reaction zone in the
laminar flame. Once the Kolmogorov scale becomes smaller, it is
hypothesized that the turbulent cascade is able to penetrate the
internal structure of the flamelets. As a result, turbulent transport
becomes comparable to, or exceeds, molecular diffusive processes, and
the flamelet width and burning velocity increase.

Overall, in this picture the actions of large and small scales are
quite distinct. Large scales determine the global turbulent burning
speed, which increases compared to the laminar burning speed in
proportion to the increase in the flamelet surface area. At the same
time, small scales affect the turbulent flame speed only by modifying
the local burning velocity of individual flamelets.

This dichotomy in the action of the large and small scales by and
large has its root in the behavior of a passive scalar in turbulent
nonreactive flows. The fact that in nonreactive turbulence the
inertial range extends to scales smaller than $\delta_L$, however,
does not automatically mean that turbulent motions of the same
intensity penetrate inside the flamelets, nor does it mean that they
have the same effect on the internal flamelet structure as they would
have on a passively advected scalar.

To date, the majority of research efforts have focused on regimes
characterized by smaller ratios of turbulent velocities to the laminar
flame speeds (see reviews by \cite{Driscoll, Bilger}), where there are
the most applications of practical interest. In these cases,
penetration by small-scale turbulent motions into the internal
flamelet structure is either completely suppressed, or it is heavily
dominated by flame wrinkling due to large-scale motions. Results of
these efforts are in agreement with the Damk\"ohler concept discussed
above, and they support the classification of such regimes as
``wrinkled'' or ``corrugated'' flamelets.

The investigation of the regimes in which turbulent velocities are
significantly larger than the laminar flame speed, $S_L$, on all
scales, including that of the laminar flame thickness, presents a
number of both experimental and numerical challenges. Hereafter, we
refer to this mode as high-speed turbulent combustion, which
encompasses regimes such as thin and broken reaction zones, as well as
distributed flames. These are the regimes in which substantial flame
broadening by turbulent transport has been hypothesized. We also
assume that turbulence is the only process that can affect the
structure and properties of the flame. We do not consider situations
in which the flame is altered or broadened by fuel preconditioning,
compression of the overall system, or propagation of large-scale
shocks.

Probing such regimes experimentally requires either high turbulent
intensities or low laminar flame speeds. Generating and sustaining
high-speed reactive turbulence with desired properties may be
difficult in a laboratory setting. More important, however, is that if
$S_L$ is high, the maximum ratios of the integral turbulent velocity
$U_l$ to $S_L$ cannot be too large, or the overall flow will no longer
be subsonic. Therefore, almost all studies that were able to achieve
$U_l/S_L \gtrsim 10 - 20$ used lean hydrocarbon fuels, which have low
values of $S_L$ \cite{Buschmann, Dinkelacker, Mansour} (also see the
review by \cite{Driscoll}).

Of particular interest is the work by Dunn et al. \cite{Dunn}, in
which a piloted premixed jet burner was used to achieve $U_l/S_L$ in
the range $40 - 390$, corresponding to Karlovitz numbers $Ka = 100 -
3500$ and Damk\"ohler numbers $Da = 0.069 - 0.0053$. In their lowest
jet-velocity case, which would be characterized as being well into the
distributed-flame regime, there was no evidence of flame broadening
based on the temperature and OH images. This is in contradiction to
predictions of the traditional classification of the combustion
regimes. Substantially higher turbulent intensities were required to
produce a flow structure that was characterized by the authors as a
distributed mode of burning, although this determination was done
mostly on qualitative grounds.

In numerical simulations, generating turbulence of arbitrary intensity
does not present any complications. Instead, modeling high-speed
turbulent reactive flows faces different challenges that can be as
restrictive as those encountered in experiments. One major difficulty
is that the Kolmogorov scale can be substantially smaller than
$\delta_L$, thereby making fully-resolved DNS difficult even with the
largest computational resources. Furthermore, when fluid velocities
reach a substantial fraction of the speed of sound, it is necessary to
use fully compressible integration algorithms for which the numerical
time step is limited by the CFL condition. This further increases the
computational cost.

Several studies of chemical flames achieved values of $U_l/S_L$ in the
range $3 - 10$ in a flat-flame configuration
\cite{Boger,Bell,Tanahashi} and $U_l/S_L = 20$ in the case of a bunsen
flame \cite{Sankaran} (also see the review by \cite{Driscoll}). In all
of these studies, the reaction zone has a thin sheet-like structure.
Moreover, Bell et al. \cite{Bell} demonstrate that the increase in
turbulent velocity can almost completely be accounted for by the
corresponding increase in the flamelet surface area, although the
former appears to be $\sim \! 10\%$ larger than the latter. Note that
the studies considering the flat-flame configuration
\cite{Boger,Bell,Tanahashi} used decaying turbulence, which precluded
the system from reaching a steady state.

Aspden et al. \cite{Aspden} reported flame broadening and formation of
a distributed reaction zone in a DNS under the action of fast
turbulence. This study considers the interaction of a thermonuclear
flame, driven by $^{12}$C-$^{12}$C reactions, with steady driven
turbulence under the conditions characteristic of the interior of
a white dwarf during the late stages of the type Ia supernova
explosion. In the fastest turbulence case, the value of $U_l/S_L
\approx 70$ was achieved corresponding to the Karlovitz number
$Ka = 228$, where $Ka = (\delta_L/L_G)^{\sfrac{1}{2}}$ and $L_G$ is
the Gibson scale defined in eq.~(\ref{e:DRLG}). This simulation showed
the formation of a broad turbulent flame with a complex temperature
and burning-rate structure.  Moreover, joint PDFs of various
quantities, such as carbon mass fraction and temperature, were
characteristic of a Lewis number $Le = 1$ system, whereas the actual
values of $Le$ are extremely large. This fact, along with wide smooth
averaged profiles of various quantities throughout the turbulent
flame, led to the conclusion that the resulting flame is dominated by
turbulent transport and presents an example of a broad distributed
reaction zone.

Even given the efforts discussed above, the physics of high-speed
turbulent combustion still has many unanswered questions. With the
exception of the experimental results of \cite{Dunn} and numerical
results of \cite{Aspden}, evidence of significant flame broadening and
the formation of distributed reaction zones is scarce \cite{Driscoll}.
Moreover, results of all studies suggest that such broadening, even if
it does occur, requires substantially higher turbulent intensities
than traditionally predicted. It remains unclear whether the effects
of turbulence are indeed so much weaker than can be expected based on
simple physical grounds and, if this is the case, what mechanisms are
responsible for suppressing the action of turbulence.

The work by \cite{Dunn} and \cite{Aspden} emphasizes the importance of
detailed and reliable diagnostics. While both studies present
interesting experimental and numerical investigations of the
previously unexplored regimes, they both rely on indirect evidence in
their determination of the observed flame broadening. Of paramount
importance are the measurements that would, for example, directly
answer questions regarding the actual multidimensional structure of
the region in which most of the energy release takes place, the nature
of flame broadening, and the effects of turbulence on different
regions of the flame.

Our goal in this work is to address several key questions concerning
the dynamics and properties of the turbulent flames formed under the
action of high-speed turbulence. First of all, it is important to
understand (1) the global properties of the turbulent flame brush in
this regime, \ie its width, speed, etc.; (2) the internal structure of
the flame brush; and (3) the internal structure of the flamelets
folded inside the flame brush, if such flamelets can be identified at
all. It is important to address the last two points quantitatively
through direct measurement. All of these characteristics are
determined by turbulence-flame interactions, and the resulting system
is the product of the collective action of the full turbulent energy
cascade. Therefore, it is also important to determine the relative
role of large- and small-scale motions on the global and local
properties of the turbulent flame.

In this paper, we begin to address these questions by presenting three
simulations, which are designed to represent the turbulent combustion
process in an unconfined (ideally infinite) space. Initially, the
flame is a planar surface separating half of the domain containing
fuel from the half of the domain containing products and immersed in a
high-intensity turbulent flow field with a Kolmogorov-type spectrum
\cite{K41}. It is assumed that the energy injection scale, as well as
the turbulent integral scale, are both finite and much smaller than
the overall system size. The evolution of such a system represents the
evolution of an infinite globally planar turbulent flame brush. Such
an idealized setting allows us to exclude effects such as walls,
boundaries, and the system geometry from consideration and to focus on
the role played by turbulence.

We consider driven turbulence of sufficiently high intensity to place
the system into the regime which is transitional, according to the
traditional classifications, between thin and broken reaction
zones. This is the regime in which turbulence would be expected to
have an effect on the structure not only of the preheat but also of
the reaction zone. By considering a stoichiometric reactive mixture,
we minimize the thermo-diffusive effects as well as the possibility of
flame extinction under the action of intense turbulence. We use a
simplified reaction model designed to accurately represent the laminar
flame properties of the stoichiometric H$_2$-air mixture. A relatively
high laminar flame speed of this fuel leads to turbulent velocities
that are a substantial fraction of the sound speed in cold fuel. The
resulting flow can no longer be considered incompressible, which
requires the use of a high-order, dimensionally unsplit, fully
compressible integration method.

The three simulations on which we focus in this paper are part of a
larger series of numerical models. Their goal is to survey the full
range of subsonic high-speed turbulent combustion regimes in a variety
of reactive mixtures. This work is intended to identify the overall
framework for future analysis of such numerical models.

\section{Formulation of the Problem}
\label{Methods}

\subsection{Physical model}
\label{PhysModel}

We solve the system of unsteady, compressible, reactive flow equations,
\begin{eqnarray}
\pd{\rho}{t} + \nabla \cdot (\rho \vect{u} )     & = & 0,
\label{Euler1} \\
\pd{\rho \vect{u}}{t} + \nabla \cdot ( \rho \vect{u} \otimes \vect{u}) 
				+ \nabla P       & = & 0,
\label{Euler2} \\
\pd{E}{t} + \nabla \cdot \big((E+P) \vect{u}\big) 
        - \nabla \cdot \big(K \nabla T\big)      & = & -\rho q \dot{w},
\label{Euler3} \\
\pd{\rho Y}{t} + \nabla \cdot \big(\rho Y \vect{u}\big)
	- \nabla \cdot \big(\rho D \nabla Y\big) & = & \rho \dot{w}.
\label{Euler4}
\end{eqnarray}
Here $\rho$ is the mass density, $\vect{u}$ is the velocity, $E$ is
the energy density, $P$ is the pressure, $Y$ is the mass fraction of
the reactant, $K$ and $D$ are the coefficients of thermal conduction
and molecular diffusion, $q$ is the chemical energy release, and
$\dot{w}$ is the reaction source term. The equation of state is that
of an ideal gas. We do not include physical viscosity in our model.
Therefore, dissipation is provided by the numerical viscosity. This
issue, and the way we use the inherent numerical viscosity in the
algorithm, will be discussed in detail in \S~\ref{Setup}. Finally,
since our method of turbulence driving, described in
\S~\ref{NumMethod}, does not involve modifying the fluid equations,
eqs.~(\ref{Euler1}) - (\ref{Euler4}) do not contain any forcing term.

Chemical reactions are modeled using the first-order Arrhenius kinetics
\begin{equation}
\frac{dY}{dt} \equiv \dot{w} = -A\rho Y \exp \Big(-\frac{Q}{RT}\Big),
\label{e:ReacModel}
\end{equation}
where $A$ is the pre-exponential factor, $Q$ is the activation energy,
and $R$ is the universal gas constant.

We assume that both thermal conduction and species diffusion have
similar temperature dependence
\beq
D = D_0 \frac{T^n}{\rho}, \quad
\frac{K}{\rho C_p} = \kappa_0 \frac{T^n}{\rho},
\label{e:CondDiff}
\eeq
where $D_0$, $\kappa_0$, and $n$ are constants, and $C_p = \gamma
R/M(\gamma-1)$ is the specific heat at constant pressure. The Lewis
number $Le = \kappa_0 / D_0$ is set to be equal to one and it is
independent of the thermodynamic conditions of the flow. External
forces, the Soret and Dufour effects, pressure gradient diffusion as
well as the effects of radiation are assumed to be negligible. The
last assumption, however, requires further verification. Contributions
of the molecular dissociation, atomic ionization, and radiative heat
losses to the equation of state are assumed to be absorbed by the
model value of the adiabatic index $\gamma$.

Reaction model parameters are summarized in Table~\ref{t:Params}. They
are based on the simplified reaction model of \cite{Gamezo} designed
to represent the stoichiometric H$_2$-air mixture. The values of the
transport coefficients are comparable to those of air at the same
conditions. This reaction model reproduces the key characteristics
both of the laminar flames and the multidimensional detonations in the
given reactant mixture, such as the laminar flame width, speed,
detonation velocity and detonation cell size, etc. as well as the
dependence of these quantities on pressure and temperature. It has
also been demonstrated to provide good qualitative agreement with
experimental data in more complex applications involving flame
acceleration and deflagration-to-detonation transition in channels
with obstacles \cite{Gamezo,Gamezo2}. These properties of the reaction
model, along with its low computational cost, make it a practical
choice for large-scale multidimensional simulations aimed at
qualitative analysis of the turbulence-flame interaction in
stoichiometric H$_2$-air mixtures and other similar fuels. It is
important to keep in mind that the first-order Arrhenius kinetics is
not able to capture the full complexity of hydrogen combustion.
Therefore, certain care must be exercised when comparing results
obtained using this model with the actual experimental data.

\subsection{Numerical method}
\label{NumMethod}

To model the process of turbulence-flame interaction, we use the code
Athena-RFX \cite{Poludnenko} -- the reactive flow extension of the
magnetohydrodynamic code Athena \cite{Athena, Gardiner,
Stone2}. Athena-RFX is a fixed-grid massively parallel code. It
implements several higher-order fully conservative Godunov-type
methods for integration of fluid equations. In this work we use the
method based on the fully unsplit corner transport upwind (CTU)
algorithm of \cite{Colella} and its three-dimensional (3D) extension
presented in \cite{Saltzman}. In particular, Athena implements a
variant of this method \cite{Gardiner}, which requires six Riemann
solutions per cell instead of the twelve in the original method of
\cite{Saltzman}. This integration scheme uses PPM spatial
reconstruction \cite{PPM} in conjunction with the approximate
nonlinear HLLC Riemann solver to achieve 3rd-order accuracy in space
and 2nd-order accuracy in time. A detailed description and an
extensive suite of tests of the hydrodynamic integration algorithm and
its implementation in Athena can be found in \cite{Athena, Gardiner}.

Diffusive transport is incorporated into the hydrodynamic integration
algorithm by calculating net diffusive fluxes across each cell along
each dimension. This is done using second-order finite differences
with flux matching to ensure that the conservation properties of the
method are preserved. Those fluxes are then added to the hydrodynamic
fluxes prior to the final update of the state vector in each
cell. Source terms, describing chemical reactions, are coupled using
Strang splitting to ensure the second-order accuracy of the overall
solution.

The overall resulting solver is formally second-order accurate and it
is capable of providing the accuracy of the planar laminar flame
solution of $\lesssim \! 1 - 2\%$ even with the resolution of $\sim \!
4$ cells per $\delta_L$. We refer to \cite{Poludnenko} for further
details of the implementation of the reactive-diffusive extensions in
Athena-RFX, as well as the results of tests including convergence
studies.

\subsection{Turbulence driving method}
\label{TurbMethod}

We investigate the process of flame interaction with steady
homogeneous isotropic turbulence, described by the classical
Kolmogorov theory \citep{K41}. In the absence of persistent energy
injection into the system at a large scale, turbulence decays on a
characteristic time scale of the large-scale eddy turnover time. Under
conditions considered later in this paper, this time scale is almost
four times shorter than the laminar flame self-crossing time
$\delta_L/S_L$ (see Tables~\ref{t:Params} and \ref{t:Props}).
Consequently, in order to study the quasi-steady regime of
turbulence-flame interaction, the flow must be constantly stirred at
the largest scale to create and maintain a steady energy cascade to
smaller scales.

We use a spectral turbulence-driving method similar to the one used in
\citep{Stone,Lemaster}. For completeness here we summarize the key
algorithmic stages of the method. More detailed discussion of the
method, along with the analysis of its properties and an extensive
suite of tests demonstrating its performance, can be found in
\cite{Poludnenko}.

As the first step, we consider Fourier transforms of the velocity
perturbations $\delta \hat{\vect{u}}'(\vect{k})$, with each component
$\delta \hat{u}'_i$ being an independent realization of a Gaussian
random field with zero mean and unit variance. Then a given isotropic
energy injection spectrum $\delta \mathcal{E}(k)$ is superimposed on
$\delta\hat{\vect{u}}'(\vect{k})$
\beq
\delta \hat{u}_i (k) =
\frac{\sqrt{\delta \mathcal{E}(k)}}{k} \delta \hat{u}'_i (k),
\label{e:u1}
\eeq
where $k = ||\vect{k}||$. In principle, the amplitude and phase of each
component $\delta \hat{u}_i$ at every point $\vect{k}$ can be
independently adjusted to produce an energy-injection pattern of
arbitrary complexity.

The computational domain has volume $V = L_1 L_2 L_3$, where $L_1
\le L_2 \le L_3$. In this work we inject energy only at the scale
$L = L_1$ to obtain the Kolmogorov-type spectrum. In particular,
\beq
\delta \mathcal{E} = \left\{
\begin{array}{l}
0 \ \ \forall \ \ \vect{k} = \{k_1,k_2,k_3\}:
k_1 = k_2 = k_3 = 0, \ \
k_i \ne \frac{\displaystyle 2\pi}{\displaystyle L_1}, \\
1 \ \ \forall \ \ \vect{k} = \{k_1,k_2,k_3\}:
k_i = 0, \ \
k_i = \frac{\displaystyle 2\pi}{\displaystyle L_1}.
\end{array}
\right.
\label{e:Einj}
\eeq
The mode with all zero components is not driven since that would
simply be inducing a uniform bulk flow through the domain.

Finally, non-solenoidal components of the velocity perturbations are
removed using the orthogonal projection operator, which ensures that
$\nabla \cdot \delta \vect{u}(\vect{x}) = 0$.

An inverse Fourier transform of $\delta \hat{\vect{u}}(\vect{k})$
gives $\delta \vect{u}(\vect{x})$, the velocity perturbation field in
the physical space. Then $\delta \vect{u}(\vect{x})$ are added to the
velocity field $\vect{u}(\vect{x})$ in the domain on every time step,
after first being normalized to ensure constant rate $\varepsilon$ of
kinetic-energy injection. Moreover, the total momentum in the
perturbation field is subtracted from $\delta \vect{u}(\vect{x})$ to
ensure that no net momentum is added to the domain, \ie $\int \rho
\delta \vect{u} = 0$. The overall perturbation pattern is regenerated
at every time interval $\Delta t_{vp} \approx 5\Delta x/c_s$, where
$c_s$ is the sound speed in the domain.

This turbulence driving method produces statistically steady,
isotropic, and homogeneous turbulent flows with energy spectra of
arbitrary complexity \cite{Poludnenko}.  In particular, it is possible
to obtain Kolmogorov-type turbulence with the inertial range of the
energy cascade extending all the way to the energy injection
scale. This is crucial, given the limited range of spatial scales
typically accessible in a simulation.  The method does not introduce
any large-scale anisotropies or global flows. Moreover, since the
velocity perturbation field is purely solenoidal, no compressions or
rarefactions are artificially induced as a result of driving. This is
particularly important in the case of the reactive flows, in which the
rate of energy generation can be very sensitive to the local
variations in temperature and pressure.

\subsection{Problem setup and summary of performed simulations}
\label{Setup}

Table~\ref{t:Runs} summarizes key parameters of the performed
simulations. The main difference between the three models is the
resolution, which progressively increases from $\Delta x = \delta_L/8$
in S1 to $\Delta x = \delta_L/32$ in S3. One of the principal goals of
this work, discussed in \S~\ref{Intro}, is to investigate the effects
of small-scale motions and, thereby, to differentiate them from the
role played by large scales. The most direct way to achieve this is by
varying the viscosity. In particular, changing the resolution changes
the numerical dissipation, and this causes the effective viscosity in
the domain to vary. This allows us to control the amount of kinetic
energy contained on scales comparable to or smaller than $\delta_L$,
while preserving the energy spectrum on larger scales.

\begin{figure}[t]
\centering
\includegraphics[clip, width=0.5\textwidth]{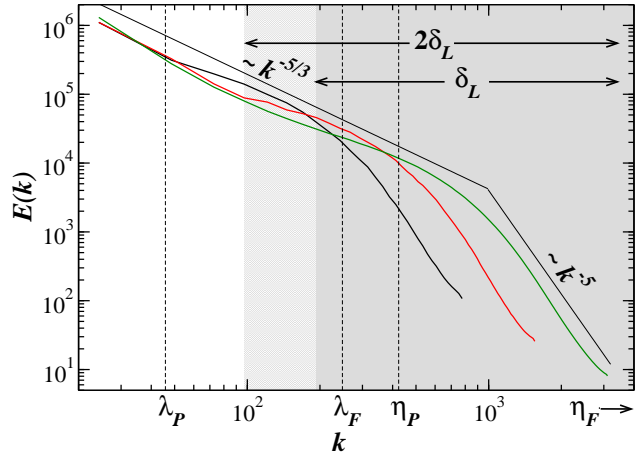}
\caption{Instantaneous energy spectra for simulations S1 (black),
S2 (red), and S3 (green) at a time immediately prior to ignition.
Shaded regions illustrate scales associated with thermal width
$\delta_L$ and full width $2\delta_L$ of the laminar flame (cf.
Fig.~\ref{f:Struct}). Vertical dashed lines show wavenumbers
corresponding to the physical Taylor microscales in the product,
$\lambda_P$, and fuel, $\lambda_F$, as well as the physical Kolmogorov
scale in the product, $\eta_P$, based on the value of the viscosity
coefficient discussed in \S~\ref{Setup}. The wavenumber corresponding
to the Kolmogorov scale in the fuel, $\eta_F = 1.18\times 10^{-3}$ cm,
is located outside the range of the graph.}
\label{f:FlameSpectra}
\end{figure}

Figure~\ref{f:FlameSpectra} shows instantaneous spectra in the domain
prior to ignition in simulations S1 -- S3. These spectra represent the
flow field in the nonreactive turbulence. While the spectra are
virtually the same on scales $\gtrsim 2\delta_L$, they vary greatly in
energy contained on small scales. In the least resolved case, S1,
motions on all scales $\lesssim \delta_L$ are affected by dissipation.
In the most resolved case, S3, the inertial range extends to scales
which, in the reactive flow, would be deep inside the flame. As a
result, energy on the small scales varies by up to a factor of 30
between these two cases. By comparing the flow structure in these
simulations, it is then possible to establish the effects of
small-scale turbulence on the turbulent flame. In particular, this
allows us to determine the degree to which turbulent cascade is able
to penetrate and alter the internal structure of the flame.

A similar analysis could be performed by including physical viscosity
in the model and by varying it parametrically. Such an analysis,
however, would not represent the behavior of an actual fuel any more
than the one which relies on numerical viscosity, unless the viscosity
parameter is close to its real physical value. Moreover, there are
several practical complications associated with using physical
viscosity instead of the numerical one. Figure~\ref{f:FlameSpectra}
shows the wavenumbers associated with the physical Taylor and
Kolmogorov scales in both the product and fuel in a stoichiometric
H$_2$-air mixture. These values were determined using the temperature
dependence of the viscosity coefficient similar to that of other
diffusion coefficients, namely $\nu = \nu_0 T^n/\rho$ where $n = 0.7$
and $\nu_0 = 2.9 \times 10^{-5}$ g/s$\cdot$cm$\cdot$K$^{0.7}$
corresponding to $Pr = Sc = 1$ \cite{Gamezo}. It can be seen that
resolving the Kolmogorov scale in the cold fuel would require
substantially higher grid resolutions. Achieving such resolution,
while maintaining a reasonably large separation between the scales $L$
and $\delta_L$, would be difficult to impossible with current
computational resources. For higher turbulent intensities, $\eta_F$
would be even smaller. In addition, physical viscosity results in a
much shallower and wider dissipation range. Given the limited range of
spatial scales accessible in a numerical simulation, this would
substantially restrict the possible extent of the inertial range in
the flow.

Figure~\ref{f:FlameSpectra} shows that in cases S1 and S3, the
numerical Taylor scales, \ie the largest scales in the domain affected
by dissipation, are similar to physical Taylor scales in product and
fuel, respectively. Therefore, the performed simulations do capture
both the smallest and the largest extents of the inertial range
present in the physical stoichiometric H$_2$-air mixture.

These considerations show that fully resolved DNS models with physical
viscosity are impractical for our study. More important, however, is
that it is possible to perform a detailed analysis of the effects of
small scales on the turbulent flame by varying numerical resolution
without resorting to fully resolved DNS simulations. In particular, if
a certain property of the turbulent flame does not change or appears
to be converged with increasing resolution, it indicates that its
evolution is determined only by large scales. On average, the flow on
large scales is independent of small scales. Therefore, if large-scale
motions are properly captured in a simulation, as is the case in the
models presented here, then the evolution of such a property cannot be
affected by the even smaller scales that would exist in a real system
due to a much smaller physical viscosity.

Methods that rely on numerical viscosity to provide kinetic-energy
dissipation in the grid, while explicitly capturing the inertial range
of the energy cascade, are often referred to as implicit large-eddy
simulations (ILES). The feasibility of this approach was first
suggested by Boris \cite{Boris} and it has since been extensively used
in turbulence modeling. In particular, a detailed analysis of the
effects of numerical viscosity in the context of the PPM method,
similar to the one used in this work, was performed by Porter et
al. \cite{Porter}. An example of similar analysis for a different
higher-order algorithm, namely flux-corrected transport, can be found
in \cite{Jun}. A comprehensive overview of this class of methods is
given in \cite{Grinstein}.

Aside from controlling the amount of energy on small scales, grid
resolution also plays its traditional role. It must be high enough to
provide sufficient accuracy of the overall solution on all scales. As
was demonstrated in \cite{Poludnenko} for Athena-RFX, a resolution of
even 8 cells per $\delta_L$ is sufficient to reproduce very accurately
properties of the planar laminar flame. {\it A priori}, it is not
clear, however, that this resolution is enough to capture the complex
dynamics of the turbulent flame, which contains highly distorted
flamelets. Strong turbulence can fold individual flame sheets with the
curvature radius comparable to $\delta_L$ or bring two flame sheets
together. This would form cusps, whose velocity of propagation can be
substantially larger than that of a planar laminar flame. Properly
capturing the rate of burning in such cusps depends sensitively on the
code's ability to accurately model thermal flux focusing in regions of
high flame curvature. Even the high-order dimensionally unsplit
algorithms introduce some degree of anisotropy to the thermal flux,
which tends to align itself with the grid. At low resolution, this can
substantially affect thermal transport in cusps. The increasing
resolution in simulations S1 -- S3 allows us to carry out a
convergence study to address these issues.

All simulations were performed in the domain with the high
length-to-width ratio of 16:1. The longest dimension of the domain is
assumed to be along the $z$-axis. Such long domains are
computationally expensive, however they are needed to be able to
follow the flame evolution for extended periods of time, namely for
$16\tau_{ed}$. We learned empirically that this extended domain size
is necessary to accommodate flame-brush motion with respect to the
grid during that time and to ensure that the flame brush remains
sufficiently far from the outflow boundaries in order to minimize
their effect on the system.

The domain in all calculations was initialized with uniform density
$\rho_0$ and temperature $T_0$ (see Table~\ref{t:Params}). In
simulations S1 and S2, initial fluid velocities were set to 0. In
contrast, in S3 the velocity field was initialized with the ideal
energy spectrum $\mathcal{E}(k) \propto k^{-\sfrac{5}{3}}$ extending
from the energy injection scale $L$ to the numerical Kolmogorov scale
$\eta = 2\Delta x$. This initial spectrum was normalized to ensure
that at $t = 0$ the total kinetic energy in the domain was equal to
its predicted steady-state value as described in
\cite{Poludnenko}. The main advantage of this method of velocity
initialization is that it allows the flow to reach its equilibrium
state much faster. In particular, the equilibration time is
$2\tau_{eq}$ in S3, whereas it is $3\tau_{eq}$ in S1 and S2 in which
initial velocities are zero \cite{Poludnenko}. This is due to the much
shorter time needed to populate all spectral modes and for the steady
energy cascade to develop. Moreover, the rate of energy dissipation is
close to its equilibrium value practically from $t = 0$, thus
preventing the build-up of excess kinetic energy and, thereby, also
shortening the equilibration phase. A detailed discussion of this
method of turbulent flow field initialization can be found in
\cite{Poludnenko}.

It is shown in \cite{Poludnenko} that in nonreactive turbulence, the
long-term system evolution is not sensitive to the form of the
velocity field in the domain at $t = 0$, provided that sufficient time
was given for the system to reach its equilibrium state. Further
results will demonstrate that this is also the case in reactive
turbulence.

Steady driving with constant rate $\varepsilon$ of energy injection
into the domain was applied to the system for the total duration of
all simulations. The energy injection scale is always assumed to be
the domain width $L$. The value of $\varepsilon$ was chosen to produce
a turbulence field of sufficiently high intensity that, however, was
low enough to minimize the probability of creating the weak transonic
shocklets that arise from intermittency in turbulent flow. In fast
turbulent flows, which can still be nominally characterized as
subsonic based, for instance, on the value of $U_{rms}$, such
shocklets can represent a substantial part of the flow field
\cite{Poludnenko}.

Table~\ref{t:Runs} lists a number of velocity characteristics along
with the integral scale of the steady-state turbulent flow prior to
the moment of ignition. In the simulations, we do not follow the
evolution of the nonreactive turbulence in the steady state long
enough to extract time-averaged values of those parameters. Therefore,
Table~\ref{t:Runs} lists theoretically predicted equilibrium values
based on the expressions given in \cite{Poludnenko}. Analysis
performed in \cite{Poludnenko} shows that such estimates generally are
accurate to a few percent. All derived quantities listed in
Table~\ref{t:Runs}, such as $\tau_{ed}$, $L_G$, $Da$, $Ma_F$, and
$Ma_P$, are also based on those values.

In our model we do not include any artificial cooling mechanisms to
compensate for the turbulent heating of the flow. Based on the
analysis given in \cite{Poludnenko}, the amount of energy injected
into the domain within one large-scale eddy turnover time $\tau_{ed}$
is $\approx \! 80\%$ of the total steady-state kinetic energy in the
nonreactive turbulent flow. The typical time step is $\sim \!
10^{-9}$ -- $10^{-8}$ s, or $\sim \! 10^{-4}\tau_{ed}$ (see
Table~\ref{t:Runs}). Therefore, the amount of energy injected into the
domain on every time step is $\lesssim 10^{-4}$ of the equilibrium
kinetic energy. The corresponding relative increase in the internal
energy and temperature due to the turbulent heating in the nonreactive
flow within $\tau_{ed}$ is
\cite{Poludnenko}
\beq
\frac{E_t - E_{t,0}}{E_{t,0}} = \frac{T - T_0}{T_0} =
\frac{\varepsilon \rho_0 \tau_{ed}}{E_{t,0}} = D_K\gamma (\gamma - 1)Ma_F^2,
\label{e:dEdT}
\eeq
where the constant $D_K = 0.5$ and $E_{t,0}$ is the thermal energy in
the domain at $t = 0$. Using values of $\gamma$ from
Table~\ref{t:Params} and $Ma_F$ from Table~\ref{t:Runs}, this increase
is $\approx \!  0.6\%$ of their initial values or, equivalently,
$\approx \! 1.8$ K over $\tau_{ed}$. Such a small amount of heating
cannot result in any significant fuel preconditioning and, thus, it
does not affect any of the conclusions.

In simulations S1 and S2, the flow field was allowed to evolve for the
time $t_{ign} = 3\tau_{ed}$, and in S3 for the time $t_{ign} =
2\tau_{ed}$, to develop the steady-state turbulent flow field. At the
time $t_{ign}$ a planar laminar flame with its front parallel to the
$x-y$ plane was imposed in the domain. The initial flame position
$z_{T,0}$, given in Table~\ref{t:Runs}, was assumed to be the location
of the $Y = 0.5$ point in the exact laminar flame solution for the
reaction model parameters and fuel conditions used. Values of $\rho$,
$P$, and $Y$ in each cell were reset to those obtained from the
laminar flame structure based on the cell-center $z$-coordinate.
Combustion product is located toward the left $z$-boundary and fuel is
located toward the right $z$-boundary. The velocity field was not
modified, thereby preserving the structure that developed during the
equilibration stage. Hereafter, for simplicity we refer to the moment
of ignition as $t = 0$.

Prior to $t_{ign}$, all domain boundaries are periodic. At $t_{ign}$,
boundary conditions (BCs) along the left and right $z$-boundaries are
switched to zero-order extrapolation. As the flow evolves, it develops
a complex pattern near the boundaries. Local turbulent motions may
have velocity vectors directed both in and out of the domain at
various points along the boundary.  Moreover, this is superimposed on
the large-scale net flow associated with the expansion of burnt
material.  Consequently, BCs must provide the possibility for the
fluid to move both in and out of the domain while adjusting to the
flow. We find that zero-order extrapolation BCs successfully manage to
accomplish this goal while preventing both any unphysical pressure
build-up in the domain and the formation of artificial large-scale
rarefaction waves at the boundaries.

This type of BCs is known to cause reflections of acoustic
perturbations. High-speed turbulent flow in the domain, however,
itself generates substantial acoustic noise and even weak transonic
shocklets. These features are substantially stronger than reflections
from the boundaries, so that it becomes nearly impossible to
distinguish such reflected perturbations from those generated by the
flow field itself. It would, nevertheless, be desirable in the future
to investigate the effects of other types of BCs on the system
evolution in order to assess its sensitivity to the prescribed BCs.

Fluid entering the domain with the large-scale inflow does not
initially have the proper Kolmogorov-type spectrum. Provided that the
turbulent flame brush is located sufficiently far from the boundaries,
the typical time for the fluid to travel from the boundary to the
flame is longer than a few $\tau_{ed}$. Therefore, this time is
sufficient for the incoming flow to be subjected to the external
driving in the domain and to develop the equilibrium spectrum.

Finally, given the simulation parameters discussed above and listed in
Table~\ref{t:Runs}, it is instructive to consider the location of the
regime, studied in this work, in the traditional combustion
diagrams. Figure~\ref{f:Diagrams} shows the two most widely used
examples of such diagrams, namely those described by Peters
\cite{Peters} and Williams \cite{Williams,Williams2}. Detailed
discussion of the diagrams including various lines delineating
different combustion regimes are given in
\cite{Peters,Williams,Williams2}.

It is important to emphasize that the laminar flame width, $l_F$, used
in the Peters' diagram, is different from the thermal flame width,
$\delta_L$, used throughout this paper as a characteristic length
scale associated with the laminar flame. In particular, $l_F$ is
defined as
\beq
l_F = \frac{D}{S_L},
\label{e:lF}
\eeq
where $D$ is the characteristic diffusion coefficient, and it is
assumed that $Le = Pr = Sc = 1$. The value of $D$, however, is not
constant and varies with temperature (cf. eq.~(\ref{e:CondDiff})).
Thus, the question arises concerning the value of $T$ that should be
used to define $l_F$. It is reasonable to take the temperature and
density in the exact laminar flame profile that correspond to $Y =
0.5$. This value of $Y$ represents the boundary between the preheat
and reaction zones. Moreover, in the vicinity of this point, $T$ has
the steepest gradient which is used to define $\delta_L$.  Then, using
the corresponding values $T = 1212$ K and $\rho = 2.1\times 10^{-4}$
g/cm$^3$ in eq.~(\ref{e:CondDiff}), we find that $l_F \approx
2\delta_L$, which is approximately equal to the full width of the
laminar flame. For consistency, we use this value of $l_F$ in both
diagrams as a characteristic scale of the laminar flame.

The Damk\"ohler and Reynolds numbers, as well as the Gibson scale are
then defined in the usual manner \cite{Peters}
\beq
Da  = \frac{lS_L}{l_FU_l}, \quad
Re  = \frac{lU_l}{l_FS_L}, \quad
L_G = l \ \Big(\frac{S_L}{U_l}\Big)^3.
\label{e:DRLG}
\eeq
Using these equations, values of $\delta_L$ and $S_L$ from
Table~\ref{t:Params}, and $l$ and $U_l$ from Table~\ref{t:Runs}, the
location of the regime considered here is shown in the combustion
diagrams (Fig.~\ref{f:Diagrams}). Simulations, presented here, fall in
the region that is transitional between thin and broken reaction
zones. In this regime, turbulent transport is expected to become
comparable to the molecular diffusive processes and, thereby, to start
affecting the internal flame structure not only in the preheat but
also in the reaction zone. Such break-up of the laminar structure of
the flamelets would mark the onset of the distributed mode of burning
(Fig.~\ref{f:Diagrams}, right panel). Note also that the Gibson scale,
$L_G$, is almost four orders of magnitude smaller than $\delta_L$
(Table~\ref{t:Runs}). This also suggests that turbulent cascade should
penetrate deep inside individual flamelets disrupting them. The
performed simulations allow us to investigate the validity of such
classification of this combustion regime.

\section{Results}
\label{Results}

\subsection{Overall evolution and global properties of the flame brush}
\label{Evolution}

Upon ignition, the planar laminar flame starts being wrinkled by
turbulent motions, and the turbulent flame brush gradually develops.
Within $\sim \! 2\tau_{ed}$, the flame brush reaches a quasi-steady
state in which its width $\delta_T$ and speed $S_T$ on average remain
constant.

Figure~\ref{f:isoV} shows the 3D structure of the turbulent flame
brush based on the distribution of the fuel mass fraction. A general
examination of the flame morphology in Fig.~\ref{f:isoV} already
suggests crucial similarities as well as differences in the system at
different resolutions. In all three simulations, the flame brush
represents a highly convolved flame with a thinner reaction zone and a
thicker preheat zone. The thin black line on the sides of the
computational domain marks the boundary between the preheat and
reaction zones, and the thin white line shows the location of the peak
reaction rate. Typically, these two lines closely follow each
other. The preheat zone, shown in green and blue, is much wider than
the reaction zone, and generally its shape does not follow that of the
reaction zone.

The flame surface, when viewed from the product side, appears
remarkably independent of resolution. In all cases, it is smooth on
smaller scales and curved on larger scales comparable to the turbulent
integral scale. In contrast, the flame surface on the fuel side is
wrinkled on progressively smaller scales with increasing resolution.
This suggests that turbulent motions on smaller scales affect the
flame structure in the preheat zone and their effect becomes less
pronounced with increasing temperature toward the reaction zone.

\begin{figure}[t]
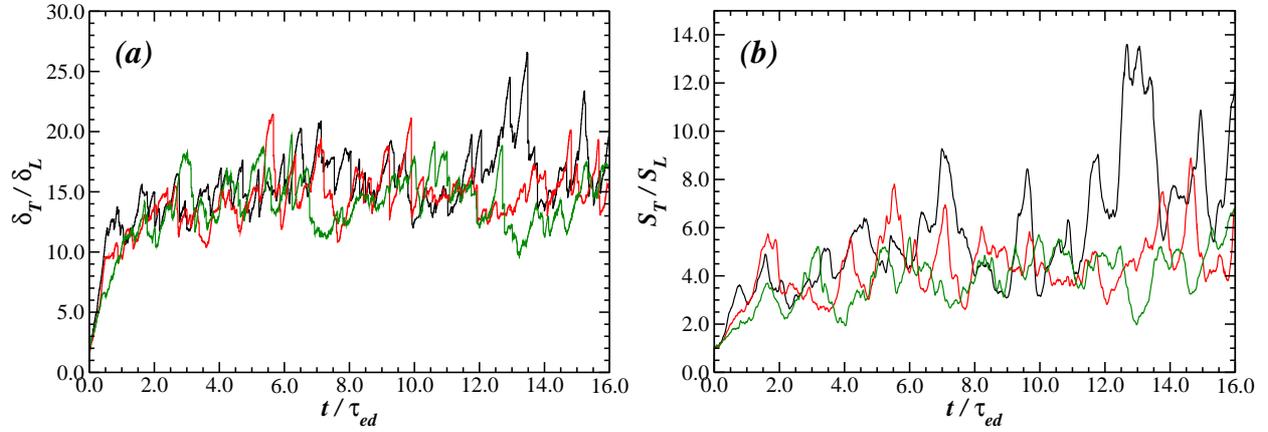

\centering 
\leavevmode
\includegraphics[clip, width=0.495\textwidth]{Lflame.v15.L8.eps}
\includegraphics[clip, width=0.495\textwidth]{Vflame.v15.L8.eps}
\caption{(a) Evolution of the turbulent flame width $\delta_T$
normalized by $\delta_L$. Note, that domain width $L = 8
\delta_L$. (b) Evolution of the turbulent flame speed $S_T$
normalized by $S_L$. In both panels, black lines correspond to
simulation S1, red to S2, and green to S3.}
\label{f:LV}
\end{figure}

Figure~\ref{f:LV} shows the evolution of $\delta_T$ and $S_T$,
normalized by $\delta_L$ and $S_L$. Hereafter, we define the width of
the turbulent flame brush as
\beq
\delta_T = z_{1,max} - z_{0,min},
\label{e:deltaT}
\eeq
where $z_{0,min}$ is such that $Y < 0.05$ for all points $(x,y,z)$
which have $z < z_{0,min}$ and $z_{1,max}$ is such that $Y > 0.95$ for
all points $(x,y,z)$ which have $z > z_{1,max}$. In other words,
$z_{0,min}$ marks the rightmost $x$-$y$-plane to the left of which is
pure product, while $z_{1,max}$ marks the leftmost $x$-$y$-plane to
the right of which is pure fuel. This is illustrated in
Fig.~\ref{f:isoS}.

There are several different ways to define the turbulent burning
velocity, as described in detail in \cite{Driscoll}. The flame brush
propagates into the fuel and its total net displacement velocity is a
function of both the global fuel-consumption speed as well as the net
fuel velocity in the domain. The net fuel velocity depends sensitively
on many factors, such as the details of the flow at the moment of
ignition, subsequent evolution of the stochastic turbulent field, and
conditions at the domain boundary. Consequently, the speed of the
flame-brush displacement with respect to the grid can vary, and it
cannot be predicted {\it a priori}.

Since we are considering an ideally infinite, unbounded system, there
are no boundaries or obstacles with respect to which the velocity of
the flame brush could be defined. At the same time, displacement
velocity with respect to the laboratory reference frame, associated
with the computational mesh, does not have physical significance. In
principle, we could associate a stationary reference frame with the
fuel located (infinitely) far from the flame brush. Such fuel, while
being in the state of turbulent motion, could be considered stationary
on average, and the displacement velocity of the flame brush with
respect to it could be defined. In practice, however, the volume of
fuel between the flame brush and the domain boundary is not
sufficiently large to allow us to perform averaging that would
eliminate spatial and temporal variability in the fuel velocity and
would provide a reliable inertial reference frame.

Therefore, we define the turbulent flame-brush velocity $S_T$ as the
global fuel consumption speed (cf. eq. (15) in
\cite{Driscoll})
\beq
S_T = \frac{\dot{m}_R}{\rho_0 L^2}.
\eeq
Here $\dot{m}_R$ is the total rate of fuel consumption inside the
flame brush, \ie the total mass of reactants which is transformed to
product per unit time. On average, the flame brush is planar in the
domain with the cross-sectional area of $L^2$. Then in steady state in
order for the flame brush to support fuel consumption rate
$\dot{m}_R$, the fuel must be supplied to the flame through the area
$L^2$ at the rate $\rho_0 S_T L^2$. Note that even though the fuel
density varies locally by a small amount due to the compressibility in
the flow, on average it remains equal to the original value $\rho_0$.
Therefore, in a perfectly steady state, $S_T$ represents both the fuel
consumption speed as well the velocity of the flame-brush displacement
with respect to stationary fuel far from the flame brush.

Table~\ref{t:Props} lists time-averaged values of $\delta_T$ and
$S_T$. Time averaging is performed over 14 large-scale eddy turnover
times. The start of the averaging interval is chosen as $t =
2\tau_{ed}$, when the system enters the quasi-steady state. While the
particular moment of the onset of this state is not precisely defined,
we find that the flame-brush parameters, discussed in this paper,
reach their time-averaged values by $2\tau_{ed}$.

Even the relatively high-speed turbulence, considered here, only
increases $S_T$ rather modestly, with the value of $S_T$ on average
saturating at $\approx 4S_L \approx 12 m/s$. The flame-brush width in
S3 is somewhat larger than the domain width, namely
$\overline{\delta_T} = 1.8 L$, and it is 7.7 times larger than the
turbulent integral scale $l$.

Table~\ref{t:Props} also lists the order of self-convergence of each
parameter. Since the computational cell size decreases progressively
by a factor of 2 between each simulation, the order of
self-convergence $O(\phi)$ of a parameter $\phi$ is defined as
\beq
O(\phi) = \log_2\Bigg(\frac{|\phi_{S1} - \phi_{S3}|}
{|\phi_{S2}-\phi_{S3}|}\Bigg).
\eeq
The table shows that both $\delta_T$ and $S_T$ converge quadratically,
as would be expected for a second-order numerical method.

In a dynamic unsteady flow, fuel consumption inside the flame brush is
not perfectly balanced by the influx of fresh fuel. Inflow of fuel can
dominate as the flame brush increases its width and the amount of fuel
inside. This is followed by a period when fuel consumption prevails,
rapidly burning the accumulated reactants and causing the flame brush
to shrink in size. While on average those two processes do remain
balanced maintaining constant average width and speed of the flame
brush, such variability is an important part of the overall
flame-brush evolution. Figure~\ref{f:LV} shows that in the lowest
resolution case S1, $\delta_T$ varies by more than a factor of two,
and $S_T$ varies by more than a factor of three. Furthermore, the
overall variability decreases with increasing resolution. This is
particularly apparent for the turbulent flame speed. Note that peak
values of $S_T$ in S3 are $\lesssim \! 6$, while in S1 they reach
$\approx \! 14$ in one episode.

Figure~\ref{f:isoV}, which shows the flame-brush structure in all
three simulations at $t = 13\tau_{ed}$, also illustrates such variable
nature of the system. Figure~\ref{f:LV} shows that at this time, S1
and S3 undergo two extreme episodes in their evolution. The flame
brush in S1 reaches its maximum width, which is $\sim 50\%$ larger
than its average value, while $S_T$ increases by more than a factor of
two. S3, on the other hand, is in the most quiescent phase with
$\delta_T$ shrinking by one third and $S_T$ being only twice the
laminar flame speed. Finally, S2 at this time is in its average state
with both $\delta_T$ and $S_T$ very close to their average values.
Therefore, aside from differences in the preheat-zone structure, the
three states of the turbulent flame shown in Fig.~\ref{f:isoV}
represent the main stages in the flame-brush evolution which are
characteristic of all three simulations. Namely, they illustrate
recurring transitions between periods of a widened flame brush,
containing highly convolved flame surface, and more quiescent stages,
when the thinner flame brush contains a smoother, flatter flame.

Qualitatively, the nature of such variability can be explained as
follows. The internal structure of the flame brush is determined by
two counteracting processes. On one hand, flame surface area is
created by turbulent motions bending and stretching the flame and,
thereby, increasing both the flame brush thickness and the flame
surface density inside the brush. On the other hand, this process is
balanced by flame-surface destruction, which is an inherently
nonlinear stabilization process preventing unbounded growth of the
flame perturbations \cite{Zeldovich1,Zeldovich2}. Flame-surface
destruction takes place in regions where the curvature radius of the
flame surface becomes comparable to the flame thickness (``cusps'') or
where the flame sheets come close to each other. Such a picture of the
balance between surface creation and destruction has been reflected in
a number of analytical and numerical models of turbulent flames (\eg
see \cite{Meneveau,Khokhlov} as well as the review by
\cite{Driscoll} and references therein).

Three distinct stages of the destruction process can be seen in the
upper and middle panels of Fig.~\ref{f:isoV}. Consider region A, where
the flame surface is tightly packed by turbulence. This results in two
flame sheets being so close to each other that their preheat zones
start to overlap. Moreover, each flamelet is also curved on scales
comparable to its thickness. Thermal flux from hot products into the
colder fuel becomes focused in such cusps, thus accelerating the
heating of reactants and effectively increasing the local burning
velocity. This makes the two flame sheets propagate faster toward each
other.

Eventually, this may lead to the configuration seen in region B. There
preheat zones of each flame sheet substantially overlap even though
the reaction zones are still separated. At this point, the whole
region acts as one large cusp with the heat flux coming from all
directions and thus rapidly heating up the unreacted material. The
temperature quickly reaches the ignition point and all of this region
becomes a reaction zone.

Examples of such merged, extended reaction zones can be seen in
regions C. Those parts of the flame quickly burn out, which reduces
the curvature and the overall surface area of the flame. Such events
are transient phenomena, and they must not be mistaken for stable
distributed reaction zones. In particular, it is very important in
experimental settings for the diagnostics to be able to properly
characterize such broad reaction zones as transients.

In fast turbulence, as present in these simulations with $Da \ll 1$,
the characteristic turbulent time $\delta_L/U_{\delta}$ at the scale
of the laminar flame width is much smaller than the laminar flame
self-crossing time $\delta_L/S_L$.  Therefore, turbulence can bring
two flame sheets together and then pull them apart before each one
propagates over the distance $\sim \! \delta_L$. As a result, flame
sheets in region A in Fig.~\ref{f:isoV} can be separated by turbulence
before they ever merge. On the other hand, the local burning velocity
in cusps can be very large, in fact, it could in principle be
infinite. Therefore, once a cusp is formed, \ie there exists a region
of significant flux focusing, it burns out on time scales short enough
for turbulence to have little or no effect on it.

In summary, periodically the flame brush undergoes a state in its
evolution when turbulence rapidly increases the flame surface which
becomes highly convolved and develops cusps. This situation is
represented by the upper panels in Fig.~\ref{f:isoV}. The local
burning velocity in cusps increases substantially to the point when
turbulent motions are no longer locally dominant. This causes cusps to
burn out quickly, leading to an increased rate of global fuel
consumption and smoothing the overall flame surface. This stage is
shown in the middle panels in Fig.~\ref{f:isoV}. Eventually, this
results in a flame brush which is substantially thinner and which
contains much less convolved flame surface, as seen in the bottom
panels of Fig.~\ref{f:isoV}. Such flatter, slower flame with fewer
cusps now again becomes susceptible to the action of turbulent
motions, which can increase its surface area, and, hence, the cycle
repeats.

Finally, Fig.~\ref{f:isoV.T} shows temperature structure in the flame
brush in S2, corresponding to the middle panels of Fig.~\ref{f:isoV}.
As expected, the temperature distribution closely follows that of the
fuel mass fraction.

\subsection{Internal structure of the flame brush}
\label{Structure}

Figure~\ref{f:isoV} suggests that even in the presence of high-speed
turbulence, the flame brush consists of highly convolved flamelets.
In order to examine this, we need to identify such flamelets and
determine their internal structure, if they are indeed present. To
accomplish this, we developed the following method.

The range of values of $Y$ and $T$ in the domain is discretized into
$n$ equal intervals. Typically, we choose $Y \in [0.01,0.99]$ and $T
\in [350.0,2135.0]$ K. The upper bound of the temperature interval is
the adiabatic flame temperature $T_P$. The lower bound is somewhat
higher than $T_0$. In the course of the simulation, fuel temperature
away from the flame slowly rises due to the turbulent energy
dissipation, thereby raising the minimum temperature value in the
domain (see \S~\ref{Setup}). The lower bound of the temperature
interval is chosen to account for the heating of the fuel and to
include only the temperature increase associated with flame
propagation. For each of the discrete values of $Y_i$ and $T_i$, with
$i$ taking values 0 to $n$, we construct an isosurface $\mathcal{S}_i$
with area $A_i$ using the ``marching cubes'' algorithm. We also
determine the volume $V_i$ bounded by the isosurfaces $\mathcal{S}_i$
and $\mathcal{S}_{i+1}$. Since the flame brush always intersects the
computational domain boundaries, this volume will always be bounded by
$\mathcal{S}_i$, $\mathcal{S}_{i+1}$, and the boundaries. Then we
construct the expression
\beq
\eta_{i+1} \equiv V_i/A_i,
\label{e:isosurf}
\eeq
assuming $\eta_0 = 0$. The $\eta_{i+1}$ gives the average distance
separating all points in the domain in which $Y = Y_{i+1}$ and $T =
T_{i+1}$ from those in which $Y = Y_i$ and $T = T_i$
respectively. Finally, $Y(\eta)$ and $T(\eta)$ give the average
flamelet structure in the flame brush. In Appendix~\ref{AppA} we
discuss in detail the properties and limitations of this method, as
well as the issues associated with its practical implementation.

For finite values of $\Delta Y \equiv Y_{i+1} - Y_i$ or $\Delta T
\equiv T_{i+1} - T_i$, the local separation between the two isosurfaces
$\mathcal{S}_i$ and $\mathcal{S}_{i+1}$ will vary from point to
point. Consider Fig.~\ref{f:isoV}. Thin white and black lines mark
isosurfaces $Y = 0.2$ and $Y = 0.6$. Regions can be identified where
they come very close to each other and where they are substantially
separated, \eg as in regions C (see also Fig.~\ref{f:isoS}). The same
is also true for temperature, as evidenced by Fig.~\ref{f:isoV.T}.  We
show in Appendix~\ref{AppA}, however, that as $\Delta Y$ and $\Delta
T$ become smaller, the variation in the local separation of the two
isosurfaces will also decrease.

Therefore, $\eta_{i+1}$, as defined in eq.~(\ref{e:isosurf}), can be
considered as a distance between consecutive isosurfaces
$\mathcal{S}_i$ and $\mathcal{S}_{i+1}$ averaged over their entire
area. Regions of large flame stretch, in which the flamelet becomes
very thin, would tend to decrease the value of $\eta$. On the other
hand, cusps, regions of significant flame folding, and broadened
reaction zones would tend to make $\eta$ larger. Thus, the flame
structure $Y(\eta)$ and $T(\eta)$, produced by this analysis for a
given instantaneous state of the flame brush, is an average of all
local realizations of the flame structure at each point of the flame
surface.

Subsequently, $Y(\eta)$ and $T(\eta)$ can be time-averaged, as a
substitute for ensemble-averaging, to obtain the statistically
dominant structure of the flamelets inside the flame brush. The
resulting structure then contains information about the statistical
weight of cusps and broadened reaction zones in the ensemble of all
possible flamelet configurations that can exist for given conditions.
Similarity of this structure to that of the planar laminar flame would
indicate that cusps and broadened reaction zones are indeed transient
phenomena, and they do not dominate the internal flame-brush
structure. On the other hand, if substantial flame broadening by
turbulent transport indeed takes place, as would be expected in the
broken reaction-zones regime, then broadened reaction zones would
dominate the statistical ensemble of possible flame configurations,
and the resulting averaged structure would be significantly wider than
the laminar flame profile.

We applied this method to the flow in simulations S1 -- S3. For S1 and
S2, $n = \delta_L/\Delta x$, as given in Table~\ref{t:Runs}, in
accordance with the criterion in eq.~(\ref{e:n}). For S3, however, $n$
was limited to a lower value, namely 16, due to the high computational
cost of reconstructing a substantially larger number of
isosurfaces\footnote{Due to the adaptive nature of the algorithm,
which implements the reconstruction method as described in
Appendix~\ref{AppA}, the actual number of isosurfaces and isovolumes
that need to be reconstructed for each variable can be significantly
larger than $n$. In particular, for $n = 32$ on average more than 100
isosurfaces need to be reconstructed for each time state with their
total number exceeding 50,000.} in the case of $n = 32$. The $Y$ and
$T$ profiles were averaged over the time interval $(2 - 16)\tau_{ed}$.
In particular, averaging was performed over 401 discrete time states
in S1, 101 in S2, and 419 in S3. The resulting profiles are shown in
Fig.~\ref{f:Struct}.

Values of the $x$-coordinate in the figure are chosen such that
$Y(x=0) = 0.5$ both in the exact laminar flame solution and in the
obtained profiles. In the exact solution, this also uniquely
determines the position of the temperature profile since there is a
precise relation between $T$ and $Y$. At the same time, the method we
are using to reconstruct the flamelet structure does not provide any
means to relate the obtained distributions of different variables, as
it is discussed in detail in Appendix~\ref{AppA}. Values of $\eta$ are
independent in each profile. It is not a common spatial coordinate,
and it simply indicates distance between consecutive isosurface values
of $Y$, $T$, etc. Each profile is a statistically averaged
representation of all existing flamelet configurations in the flame
brush. Thus, any relation between the $Y$ and $T$ profiles can only
exist in the statistical sense, namely such relation would only
indicate the most likely values of $T$ for given values of $Y$ and
vice versa. Therefore, here for simplicity we offset temperature
profiles so that $T(x=0)$ would correspond to $(T - T_0)/(T_P - T_0) =
0.5$, as in the exact laminar flame solution.

The principal conclusion that emerges from Fig.~\ref{f:Struct} is
that, on average, the turbulent flame brush interacting with
high-speed turbulence represents highly convolved flamelets that have
internal structure of a laminar flame with a somewhat broadened
preheat zone. The $Y$ and $T$ profiles are very close to the exact
laminar solution in the reaction zone. They begin to diverge from the
laminar profiles only at the outer boundary of the reaction zone,
around $x/\delta_L \approx 0.1 - 0.2$. In fact, the $Y$ profile for S3
is virtually indistinguishable from the laminar solution inside the
reaction zone. The remarkable similarity of the flamelet structure
inside the reaction zone to that of the planar laminar flame is a
partial justification for our choice of the $x$-coordinate origin for
the temperature profile.

The deviation of the $Y$ and $T$ profiles from the laminar solution
increases with distance from the reaction zone. Even for such strong
turbulence, however, this difference is fairly small, with the total
preheat zone width being increased by less than a factor of two.

The profiles, obtained for all three simulations, are very close to
each other for all values of both $Y$ and $T$. Given the statistical
nature of the distributions, this serves as evidence that the
simulations can be considered converged. More importantly, the
agreement between profiles in the preheat zone shows that thermal
conduction is enhanced predominantly by turbulent motions on scales
$\sim \! \delta_L$. The amount of energy contained on this scale is
the same in all simulations (see Fig.~\ref{f:FlameSpectra}). On the
other hand, the energy contained on smaller scales, and, thus, the
velocity, increases substantially from case S1 to S3. If those scales
were to contribute noticeably to the overall thermal transport, such
increase in energy content on those scales would be reflected in a
broader preheat zone. There is, however, no evidence of this.

Finally, an important question concerns the degree of variability of
instantaneous distributions of $Y$ and $T$ in the course of system
evolution. In particular, this question is crucial for determining the
statistical significance of the observed preheat zone broadening in
the time-averaged flame structure.

It was discussed above that $\eta$ in the distributions $Y(\eta)$ and
$T(\eta)$ is simply the distance between two consecutive isosurface
values rather than a spatial coordinate with respect to some fixed
reference frame. Therefore, in order to compare instantaneous profiles
of $Y(\eta)$ or $T(\eta)$, all values of $\eta$ must be referred to
some common point in which all profiles will, thereby, coincide. The
choice of such reference point is arbitrary and while it does not
affect the shape of the resulting time-averaged distribution, it
changes the range of variations of individual profiles. Overall, such
variations must be considered only in the vicinity of the chosen
reference point. Any difference in the shape of the profiles near this
point will cause the separation between profiles to increase with
distance from it, thereby, magnifying the perceived variability in the
profile shape.

In the case of S3, such variability range is shown in
Figs.~\ref{f:Struct}-\ref{f:Variability} with profiles for all
individual time states contained within the shaded gray areas.  All
profiles were chosen to coincide at the boundary between the reaction
and preheat zones (Fig.~\ref{f:Struct}, $Y = 0.5$ and $(T - T_0)/(T_P
- T_0) = 0.5$), at the point of peak reaction rate
(Fig.~\ref{f:Variability}a, $Y = 0.15$ and $(T - T_0)/(T_P - T_0) =
0.85$), and in the preheat zone (Fig.~\ref{f:Variability}b, $Y = 0.85$
and $(T - T_0)/(T_P - T_0) = 0.15$).

Overall, there is extremely small variability in the flame structure
both in the reaction zone (Fig.~\ref{f:Variability}a) and in the
steepest region of the profiles (Fig.~\ref{f:Struct}) throughout the
time interval used for averaging. This supports our choice of this
interval as a period of quasi-steady state in the system
evolution. There is, however, substantially more variability in the
preheat zone (Fig.~\ref{f:Variability}b). Moreover, variability of the
fuel mass fraction there appears to be somewhat larger than that of
the temperature.

Figure~\ref{f:Variability}b shows that in the preheat zone the laminar
solutions marginally lie outside the variability range of the
profiles. This confirms that the observed broadening is statistically
significant, and such flamelet structure is prevalent throughout the
evolution of the system. On the other hand, the fact that the laminar
solution is so close to the boundary of the variability range shows
how weak this effect is, even in the case of such strong turbulence as
is present in these simulations.

\section{Discussion and Conclusions}
\label{Discussion}

We have explored the dynamics and properties of a turbulent flame
formed in the presence of driven subsonic, high-speed, homogeneous,
isotropic Kolmogorov-type turbulence in an unconfined system, \ie in
the absence of walls and boundaries. The Damk\"ohler number for the
turbulent flow is $Da = 0.05$ while the Gibson scale is $L_G = 2.96
\times 10^{-4}
\delta_L$.

Numerical modeling was performed using the massively parallel, fully
compressible, higher-order, dimensionally unsplit, reactive-flow code
Athena-RFX \cite{Poludnenko}. The highest-resolution calculation S3
required approximately 100,000 CPU hours on 1024 processor cores of
the Ranger platform at TACC, and a similar amount of computational
time was necessary for data processing and reconstruction of the
flamelet structure.

\subsection{General properties of the turbulent flame}

Our results show that the turbulence-flame interaction leads to the
formation of a steadily propagating turbulent flame brush. We followed
the system evolution in this quasi-steady state for 14 large-scale
eddy turnover times, which was sufficient to provide long-term
statistics. On average, both the flame width and speed remain
constant, even though at each moment in time they can vary
substantially. We observed an increase in the time-averaged turbulent
flame speed, $\overline{S_T}$, compared to the laminar flame speed,
$S_L$, by a factor of 4. The time-averaged turbulent flame width,
$\overline{\delta_T}$, was almost twice the domain width and the
energy injection scale, $L$. It was also almost 8 times larger than
the turbulent integral scale, $l$.

An important question concerns the parameters which predominantly
determine $\delta_T$. In our simulations, the system size, which is
set by the domain width, is equal to $L$, and, thus, it is close to
$l$. It is, therefore, impossible to separate the individual role
played by the system size, $L$, and $l$ in defining the equilibrium
value of $\delta_T$. In particular, given fixed values of $L$ and $l$,
it is not clear whether a substantial increase in system size would
result in a change in $\delta_T$. Moreover, it is not clear whether in
such a large system there would still exist the observed steady flame
propagation with a fixed width that is much smaller than the system
size. Finally, an important question concerns the relation between $L$
and $\delta_T$. Does an increase in $L$ lead to a corresponding
increase in $\delta_T$, or is there a limiting value of $\delta_T$?

The study by Aspden et al. \cite{Aspden}, which used faster turbulence
relative to $S_L$ than this work, also observed a turbulent flame
propagating with a steady width and speed. In that work energy as well
was injected at the scale close to the domain width. In their
calculation that had the highest ratio of the turbulent velocity to
$S_L$, the domain size relative to the laminar flame width,
$\delta_L$, however, was almost 8 times larger than in our case,
namely $L \approx \! 65\delta_L$ vs. $L = 8\delta_L$. Yet that
simulation produced $\delta_T \approx \! 3L$, which is close to that
obtained in our simulations, namely $\delta_T \approx \!2L$. This
suggests that the width of the flame brush increases with the increase
in the driving scale and system size. Their work, however, as well
does not address the question of the individual role of these two
scales.

In systems which are much larger than the driving scale, most likely
both $L$ and $l$, rather than the domain size, play the primary role
in setting the equilibrium width of the flame brush. Indeed, the
turbulent integral scale along with the driving scale are the
characteristic scales of coherent turbulent motions. Moreover, on
scales $\lambda \gg L$, the turbulent energy spectrum has the $k^4$
dependence \cite{Davidson}. Therefore, the energy contained in those
large scales decreases rapidly with increasing scale, which minimizes
their potential role in defining the turbulent flame width. At the
same time, in situations when the domain width is close to $L$, domain
boundaries imprint an artificial periodicity on the flow. This can
constrain the growth of $\delta_T$ and, thus, set its equilibrium
value.

The main reason for the difficulty with addressing these issues is the
limited range of spatial scales presently accessible even in the most
heroic numerical calculations. The need to provide sufficiently high
resolution at the scale of the laminar flame, and still maintain some
separation between $L$ and $\delta_L$, does not allow the system size
to be more than an order of magnitude larger than $L$. This is
insufficient to provide a conclusive answer to the questions mentioned
above.

\subsection{Internal structure of the flame brush and the role of
small-scale turbulence}

A major conclusion of the simulations presented in this paper is that
in the presence of high-speed turbulence the flame brush is comprised
of highly convolved flamelets. The time-averaged internal structure of
those flamelets is very close to that of the planar laminar flame,
and, in fact, both structures are virtually identical in the reaction
zone. The preheat zone, however, does show evidence of broadening,
although this effect, while statistically significant, is fairly small
with the width of the preheat zone increasing by less than a factor of
two.

The internal structure of flamelets inside the flame brush was
reconstructed directly by a method based on the isosurfaces of the 3D
distribution of a scalar quantity, such as fuel mass fraction, $Y$, or
temperature, $T$. The overall method does not make {\it a priori} any
assumptions concerning the possible underlying flame structure or even
the presence of the flame itself.

The absence of any broadening in the reaction zone indicates that
combustion takes place in the thin reaction zone regime. Turbulent
velocities on the scale of the laminar flame width are 15 times larger
than the laminar flame speed. In principle, this would suggest that
the turbulent transport should overwhelm the molecular thermal
conduction and species diffusion.  It appears, however, that
disrupting the flame would require substantially higher turbulent
intensities. This result is in general agreement with the behavior
observed in \cite{Dunn,Aspden}.

According to the traditional combustion-regime diagrams, shown in
Fig.~\ref{f:Diagrams}, some broadening of the reaction zone would
formally be expected in the regime considered here. For instance,
according to the Williams' diagram (Fig.~\ref{f:Diagrams}, right
panel) the simulations are in the broken reaction-zone regime, while
according to the Peters' diagram (Fig.~\ref{f:Diagrams}, left panel),
they fall almost exactly at the boundary of thin and broken reaction
zones. The $Ka_{\delta} = 1$ line in the Peters' diagram is typically
constructed based on the assumption that the reaction-zone width is
$1/10$ of the full flame width. In our case such estimate is too
conservative with this fraction being $\sim \!  1/3 - 1/2$
(cf. Fig.~\ref{f:Struct}). This would lower the $Ka_{\delta} = 1$
line, thereby, also placing simulations S1 - S3 in the broken
reaction-zone regime. Of course, given the approximate nature of such
diagrams, boundaries between various combustion regimes are not really
precisely defined and, instead, they should rather be viewed as the
transition areas. Nonetheless, the fact that our results do not show
any evidence of the reaction-zone broadening suggests that the range
of the regimes in which flamelets exist may be substantially larger,
even when the uncertainties in the precise location of the indicated
regime boundaries are taken into account.

The obtained flamelet structure represents the statistically dominant
state of flamelets inside the flame brush. At the same time,
substantially broadened reaction zones can arise in the flow as a
result of collisions and mergers of flame sheets or in the regions of
high flame curvature (cusps). Such broadened reaction zones, however,
are transient in nature being the result of the rapidly changing 3D
configuration of the flame surface.  They should not be considered as
evidence of a distributed mode of burning since they are in no way
supported by turbulent transport.

The progressive increase in resolution in calculations S1 - S3 results
in a substantial increase in the energy of turbulent motions on scales
$\lambda \lesssim \delta_L$ and causes the energy cascade in
nonreactive turbulence to extend to much smaller scales. This,
however, is insufficient to disrupt the internal structure of the
reaction zone, which is demonstrated by the fact that the $Y$ and $T$
profiles, describing the internal structure of the flamelets in all
three simulations, are very similar for all values of $Y$ and
$T$. Furthermore, the structure of the broadened preheat zone does not
change between S1 and S3 suggesting that such broadening is determined
by scales $\lambda \gtrsim \delta_L$ rather than small-scale
motions. The effect of small-scale turbulence appears only in the
progressively finer wrinkling of the flame surface on the fuel side in
the coldest part of the preheat zone.

These results lead to the following important conclusion. The
turbulent energy cascade fails to penetrate the internal structure of
the flame on scales much smaller than $\delta_L$, even in the presence
of high-speed turbulence considered here. This shows that the response
of the flame to the action of turbulence is qualitatively different
from that of a passively advected scalar, which can generally be a
source of only limited insight into the behavior of the turbulent
flames.

This also suggests that the traditional definition of the Gibson
scale, $L_G$, cannot serve as a useful indicator of the efficiency
with which small-scale turbulence penetrates the flame. In particular,
in our simulations $L_G$ is almost four orders of magnitude smaller
than $\delta_L$ (see Table~\ref{t:Runs}) which, however, is not
reflected in any appreciable effects of small-scale turbulence on the
flame.

What mechanisms are responsible for suppressing the effects of
small-scale motions in the reactive turbulence? The dramatic
difference in the degree of wrinkling of the flame surface on the fuel
and product sides, observed in the highest resolution case S3,
suggests that this suppression of small-scale turbulence occurs as the
flow passes through the individual flamelets rather than the whole
flame brush. Two possible processes, which agree with this
observation, were previously suggested (see discussion in
\cite{Driscoll} and references therein). On one hand, as fluid
passes through the flame, it heats up and expands. Since the
circulation of individual eddies must be conserved during this
expansion, the rotational velocity of eddies must decrease. Moreover,
this will also shift energy from smaller to larger scales. On the
other hand, such expansion causes rapid acceleration of the fluid in
the reference frame of the flame brush. As a result, the residence
time of individual eddies inside the flame brush decreases as the
fluid quickly leaves the flame brush. It is not clear, however,
whether these two mechanisms are sufficient to completely account for
the observed behavior. It is also not known whether their effect on
small and large scales is similar or not. Answering these questions
is important for our understanding of the
turbulence-flame interaction.

Ultimately, our understanding of the turbulent combustion process is
measured, to a large extent, by our ability to predict the turbulent
burning velocity. According to the Damk\"ohler concept
\cite{Damkohler} discussed in \S~\ref{Intro}, two processes can
determine the turbulent flame speed: increase in the flame surface
area by turbulent stretching and folding of the flame, and increase in
the local flame speed by turbulent diffusion.

Similarity of the internal flame structure to that of the planar
laminar flame shows that the local flame speed cannot be substantially
increased by turbulence. The flame structure is determined by the
balance of the reactions and the diffusive processes. In our model,
the reaction rate is determined purely by the local thermodynamic
state of the fluid and, thereby, it is not directly affected by
turbulence which can only enhance the diffusive transport. For a given
one-step Arrhenius reaction model, the coefficients of thermal
conduction and species diffusion uniquely determine both the flame
structure and speed. If turbulence does indeed increase the effective
diffusive coefficients, this would increase the local flame speed, but
this would also change the flame structure. We do not find any
evidence of this.

This raises the question whether the observed turbulent flame speed,
shown in Fig.~\ref{f:LV}b, can be fully accounted for only by the
increase in the flame surface area or whether other processes
contribute to it. We will discuss this issue in detail in a separate
paper \cite{Poludnenko2}.

Finally, it is important to emphasize that averaged distributions of
various quantities, such as $Y$, $T$, or the reaction rate, $R$,
through the turbulent flame do not provide information about the
internal structure of the flame brush, and they cannot be viewed as
evidence of flame broadening. Consider Fig.~\ref{f:AverProf}, which
shows the $x$-$y$-averaged profiles of $Y$, $T$, and $R$ along the
$z$-axis in simulation S3. All profiles, including that of the
reaction rate, are smooth and the overall structure is similar to the
one shown in Fig.~\ref{f:Struct}, with the key exception that the
width of the profiles is substantially larger by a factor $\approx \!
4$. The broad, smooth profile of the reaction rate can lead to the
erroneous conclusion that this system represents a distributed mode of
burning with a substantially broadened reaction zone. If the flame
brush volume is large enough, then the average profiles of $Y$, $T$,
etc., will always be smooth and wide, since they are simply the result
of averaging many individual flamelet structures. Moreover, the
turbulent flame is the region in which $Y$ and $T$ transition from
their values in the fuel to those in the product. Therefore, those
profiles smoothly and monotonically change from their fuel to product
values, and the averaged distribution of $R$ simply varies
accordingly.

\subsection{Convergence and resolution}

What do our results say about the minimum resolution needed to provide
a converged solution in simulations of high-speed turbulent
combustion? Based on Table~\ref{t:Props}, time-averaged values of
$\delta_T/\delta_L$ and $S_T/S_L$ converge quadratically. The values
of $\overline{\delta_T}/\delta_L$ vary only by $\approx \!  3\%$ in
the two highest-resolution simulations, S2 and S3, while
$\overline{S_T}/S_L$ varies by $10\%$. Profiles of $Y$ and $T$, which
represent the internal flamelet structure, also demonstrate excellent
convergence.

The only exception to this converged behavior is the degree of
flame-surface wrinkling on the fuel side. It was discussed in
\S~\ref{Setup} that simulations with different computational cell
sizes do not represent the same physical system. The spectral energy
distribution in the domain changes with increasing resolution because
the amount of energy contained on smaller scales grows. This causes
isosurfaces in the coldest regions of the preheat zone to be wrinkled
on progressively smaller scales (see Fig.~\ref{f:isoV}). In this
context, convergence of the full system to a unique solution in the
purely numerical sense would not be expected in such turbulent
flows. At the same time, the observed convergence of all key
characteristics of the turbulent flame simply means that such small
scale motions in cold fuel do not affect the evolution of the flame
brush.

Based on these observations, we conclude that a minimum resolution of
16 cells per $\delta_L$ is required to capture the evolution of the
turbulent flame adequately. Lower resolutions tend to exaggerate both
the values and the degree of variability of all major quantities,
primarily the turbulent flame width and speed.

A minimum resolution of 16 cells is substantially higher than the
resolution of 4 cells per $\delta_L$ that is required to reproduce
with high accuracy the behavior of the planar laminar flame
\cite{Poludnenko}. This suggests that the need for higher resolution
arises due to the turbulent nature of the flow. Our results have shown
that small-scale turbulence fails to penetrate the flame and, thereby,
to modify both its internal structure and local speed. Indeed, even
the resolution of 8 cells per $\delta_L$ provided the converged
flamelet structure. Therefore, it is rather the multidimensional
effects of large-scale turbulence and, in particular, the resulting
highly curved geometry of individual flame sheets, which require
higher resolution.

In the presence of high-speed turbulence, the flame is often folded
with a curvature radius comparable to the laminar flame width (see
Fig.~\ref{f:isoV}). In such cusps, the Cartesian mesh invariably
introduces some degree of unphysical anisotropy to the thermal
flux. This effect is not present in the case of a planar laminar
flame. In the turbulent flow, however, it causes errors which at low
resolutions can become significant in comparison with the overall
solution errors. Consequently, higher resolution is necessary to
minimize this effect. In order to relax this minimum-resolution
threshold and bring it closer in line with that required for a planar
laminar flame, better multidimensional coupling of diffusive fluxes
and their improved coupling with the hydrodynamic fluxes are
needed. This is the subject for further work.

\subsection{Implications for an actual stoichiometric H$_2$-air
mixture}

The physical model used in this work is based on the first-order
Arrhenius kinetics and a simplified reaction-diffusion model of
\cite{Gamezo} for stoichiometric H$_2$-air mixture. The low
computational cost of such approach makes the large-scale
multidimensional simulations, such as the ones presented here,
computationally feasible. At the same time it also allows us to
reproduce accurately the key characteristics of laminar flames in
stoichiometric H$_2$-air, such as the laminar flame width and speed,
as well as the dependence of these quantities on pressure and
temperature.

In order to investigate the role of small-scale turbulence on the
flame, we did not include physical viscosity in our model and instead
relied on numerical viscosity to provide kinetic-energy dissipation.
As a result of this, however, our simulations, strictly speaking, do
not describe the behavior of an actual stoichiometric H$_2$-air
mixture. In light of this, what can our results say about high-speed
turbulent combustion of the actual H$_2$-air?

We have shown that scales $\lambda \ll \delta_L$ do not play a
pronounced role in determining the dynamics and properties of the
turbulent flame brush. This suggests that turbulent motions on yet
even smaller scales, which would be present in the cold H$_2$-air
mixture, should not be able to change the results either qualitatively
or quantitatively. At the same time, the evolution of the turbulent
flame is completely determined by scales $\lambda \gtrsim
\delta_L$. In all our simulations, the inertial range extends to
scales smaller than the actual Taylor scale in the product, and it
extends to scales smaller than the actual Taylor scale in the fuel in
the highest resolution case S3. Therefore, motions on all large scales
are properly captured. This fact, along with the negligible role of
small scales, suggests that our models can be viewed as representative
of the actual behavior of stoichiometric H$_2$-air.

\begin{figure}[!t]
\includegraphics[width=1.0\textwidth]{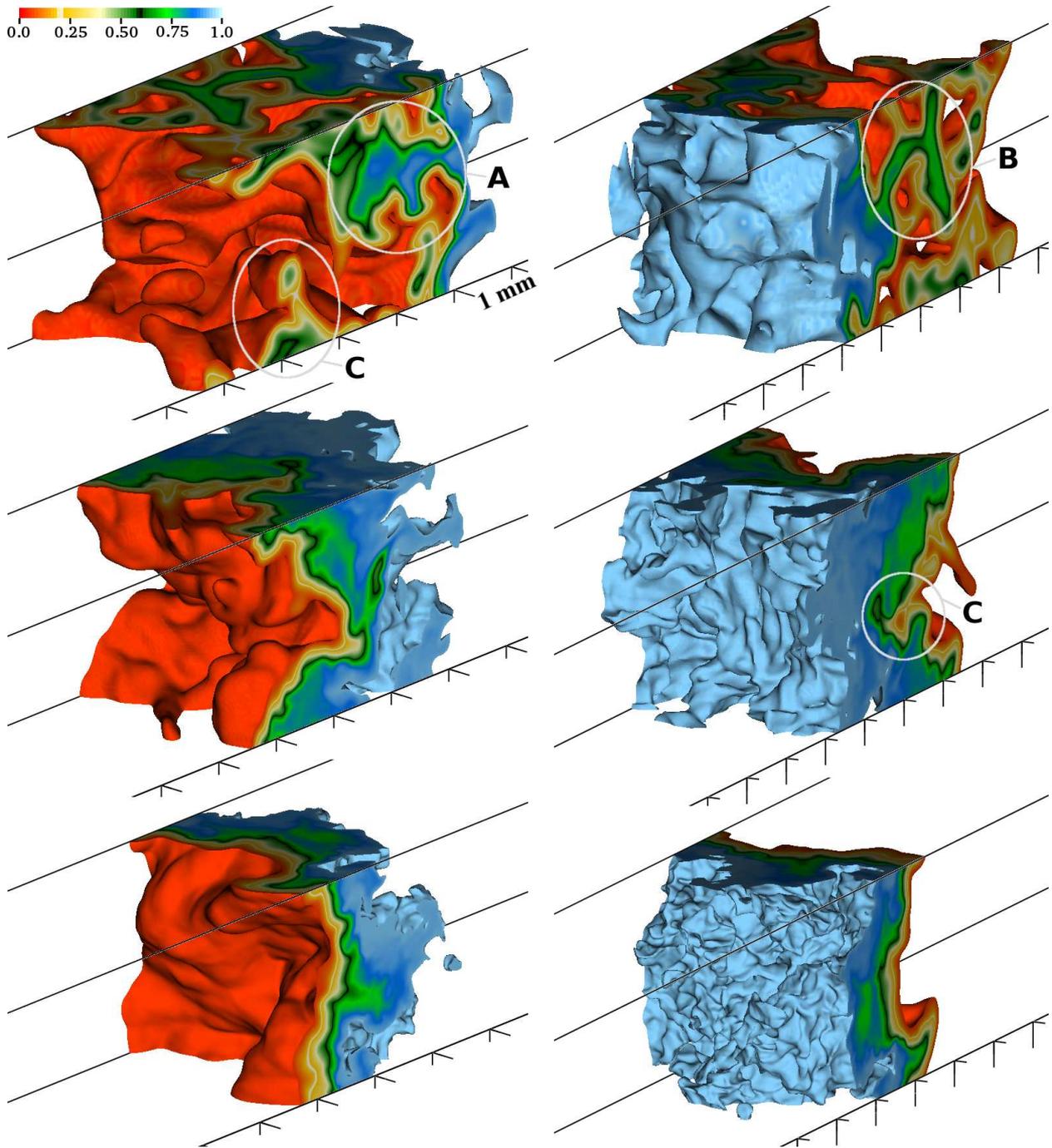}
\caption{Flame brush structure in simulations S1 (top row), S2 (middle
row), and S3 (bottom row). Shown is fuel mass fraction at $t =
13\tau_{ed}$. Left panels show view from the product side, right
panels show view from the fuel side. Bounding isosurfaces represent $Y
= 0.05$ and $Y = 0.95$. The thin black line, corresponding to $Y =
0.6$, marks the boundary between the preheat and reaction zones. The
thin white line, corresponding to $Y = 0.2$, shows the location of the
peak reaction rate (cf. Fig.~\ref{f:Struct}). Regions A, B, and C are
discussed in \S~\ref{Evolution}.}
\label{f:isoV}
\end{figure}

The only exception concerns the wrinkling scale of the flame surface
on the fuel side. The presence of turbulent motions on much smaller
scales would result in substantially finer wrinkling than shown in
Fig.~\ref{f:isoV}, even in our highest-resolution simulation. At the
same time, the flame surface on the product side is representative of
that in the actual reactive mixture.

\begin{figure}[t]
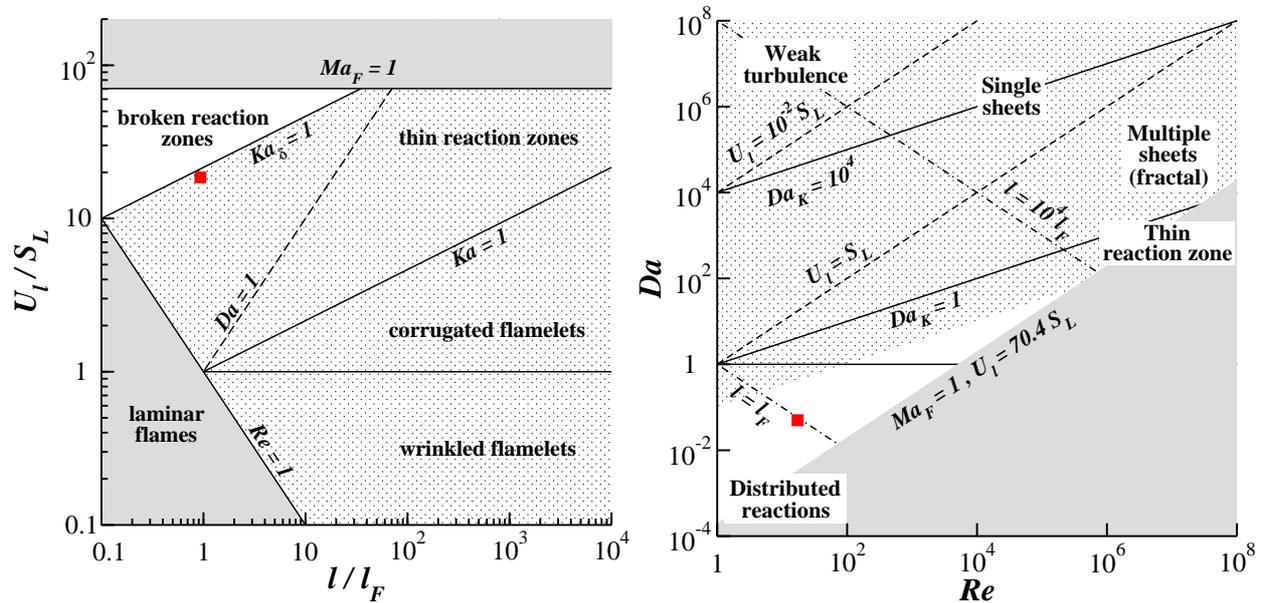

\centering 
\leavevmode
\includegraphics[clip, width=0.495\textwidth]{Peters_Diagram.eps}
\includegraphics[clip, width=0.495\textwidth]{Williams_Diagram.eps}
\caption{Combustion regime diagrams according to \cite{Peters} (left)
and \cite{Williams} (right). Red square corresponds to the simulations
presented in this work. Light shaded area shows the regime range in
which the existence of flamelets is suggested. The traditional form of
the diagrams was modified by adding the $Ma_F = 1$ line indicating the
region of supersonic turbulence in the cold H$_2$-air fuel under the
atmospheric conditions. See text for further details.}
\label{f:Diagrams}
\end{figure}

On the combustion diagrams in Fig.~\ref{f:Diagrams} we show the $Ma_F
= 1$ line, which separates the regions of subsonic and supersonic
turbulence in the cold H$_2$-air fuel under atmospheric conditions.
Even under the traditional classification, the potential range of the
regimes in which broken or distributed reaction zones would be
expected is fairly small. In presented simulations, the turbulent Mach
number in cold fuel is $Ma_{F} = 0.25$. Therefore, in order for the
turbulence-flame interaction to remain in the subsonic regime, the
turbulent velocity can be increased only moderately. This does not
appear to be sufficient, as our results, as well as the results of
other studies \cite{Dunn,Aspden}, show that substantially higher
turbulent intensities would be required to disrupt the internal flame
structure.

While faster turbulence would be more likely to affect the internal
flame structure, it would also cause compressibility effects and
turbulent heating of the fuel to become much more pronounced.
Resulting higher pressures and temperatures would increase the laminar
flame speed while decreasing its width, thereby reducing the
broadening effect of turbulence. It is also not clear what the effects
of the stronger compressive component of the velocity field in such
fast turbulence would be. In particular, local compressions and
rarefactions can also modify or disrupt the local flame structure.
Such processes would be completely distinct from the traditionally
considered action of the vortical motions associated with the purely
solenoidal part of the turbulent field.

Based on these considerations, the results presented here suggest that
the traditional classifications of combustion regimes of
stoichiometric H$_2$-air on the basis of the internal flamelet
structure and, in particular, the degree of the reaction-zone
broadening, do not accurately reflect the actual process of
turbulence-flame interaction. In particular, in all subsonic regimes,
substantial flame broadening by turbulence appears unlikely, which
suggests that the only two possible modes of turbulent stoichiometric
hydrogen combustion are the corrugated flamelets and thin reaction
zones.

The simplified reaction-diffusion model used in this work cannot
capture the full complexity of hydrogen combustion. For instance, the
Lewis number of the actual stoichiometric H$_2$-air is less than unity
which leads to cellular flames. It will be important for future work
to investigate potential effects of the more detailed description of
chemical and diffusive processes. This is the subject for future work.

In realistic experimental settings, various aspects of the high-speed
turbulence-flame interaction, discussed above, can be further
confounded by the effects of the system geometry and the potentially
complex nature of the large-scale fluid flow. It is known that
turbulent-flame evolution is not ``geometry-independent''
\cite{Driscoll}. The highly anisotropic and inhomogeneous turbulence
which arises in the presence of strong shear, counter, or jet flows
can modify the structure and properties of the turbulent flame. The
investigation of the implications of such more complex flows is the
subject for a separate study.

We conclude with the question about the universality of the observed
behavior. Is it possible to identify a set of parameters which, for
the given reactive mixture and flow conditions, would uniquely specify
the degree of flame broadening and the role of small-scale turbulence?
Are the given ratios of the turbulent intensity and the system size to
the laminar flame speed and width, which were used in our simulations,
sufficient to apply the obtained results to other fuels?

Our results appear to be in general agreement both with experimental
and numerical studies \cite{Dunn,Aspden} of high-speed turbulent
combustion in such different systems as the flat thermonuclear flame
in degenerate matter and the jet-burner flame in compressed natural
gas. In particular, it appears that, for a broad class of reactive
mixtures, the effects of turbulence on the internal flame structure
are significantly suppressed, and disrupting the flame requires
substantially higher turbulent intensities than can be predicted based
on a traditional analysis.

\vspace{5mm}

\noindent{\bf Acknowledgments} We are deeply grateful to Vadim
Gamezo for numerous valuable inputs in the course of this work. We
also thank Craig Wheeler, Forman Williams, James Driscoll, Matthias
Ihme, and Ken Bray for stimulating discussions. AYP is particularly
grateful to Tom Gardiner for his invaluable assistance with Athena and
for his continuing support. AYP also acknowledges the tireless work
and invaluable help of the VisIt developers team and especially Hank
Childs. This work was supported by the National Research Council
Research Associateship Programs in cooperation with the Naval Research
Laboratory, by the Office of Naval Research, by the Air Force Office
of Scientific Research under the grant F1ATA09114G005, and by the
National Science Foundation through TeraGrid resources provided by
NCSA and TACC under the grant TG-AST080006N. Additional computing
facilities were provided by the Department of Defense High Performance
Computing Modernization Program.

\clearpage
\begin{singlespace}

\begin{deluxetable}{ccc}
\tablecaption{Input Model Parameters and Computed Laminar Flame Properties
\label{t:Params}}
\tabletypesize{\small}
\tablewidth{0pt}
\tableheadempty
\startdata
$T_0$        &   $293$ K                                                  &   Initial temperature               \\
$P_0$        &   $1.01 \times 10^6$ erg/cm$^3$                            &   Initial pressure                  \\
$\rho_0$     &   $8.73 \times 10^{-4}$ g/cm$^3$                           &   Initial density                   \\
$\gamma$     &   $1.17$                                                   &   Adiabatic index                   \\
$M$          &   $21$ g/mol                                               &   Molecular weight                  \\
	     &                                                            &                                     \\
$A$          &   $6.85 \times 10^{12}$ cm$^3$/g$\cdot$s                   &   Pre-exponential factor            \\
$Q$          &   $46.37$ RT$_0$                                           &   Activation energy                 \\
$q$          &   $43.28$ RT$_0$ / M                                       &   Chemical energy release           \\
$\kappa_0$   &   $2.9 \times 10^{-5}$ g/s$\cdot$cm$\cdot$K$^\textrm{n}$   &   Thermal conduction coefficient    \\
$D_0$        &   $2.9 \times 10^{-5}$ g/s$\cdot$cm$\cdot$K$^\textrm{n}$   &   Molecular diffusion coefficient   \\
$n$          &   $0.7$                                                    &   Temperature exponent              \\
	     &                                                            &                                     \\
$T_P$        &   $2135$ K                                                 &   Post-flame temperature            \\
$\rho_P$     &   $1.2 \times 10^{-4}$ g/cm$^3$                            &   Post-flame density                \\
$\delta_L$   &   $0.032$ cm                                               &   Laminar flame thermal width       \\
$S_L$        &   $302$ cm/s                                               &   Laminar flame speed
\enddata
\end{deluxetable}


\begin{deluxetable}{lcccc}
\tablecaption{Parameters of Simulations\tablenotemark{a}
\label{t:Runs}}
\tabletypesize{\small}
\tablewidth{0pt}
\tablehead{
\colhead{}             &
\colhead{S1}           &
\colhead{S2}           &
\colhead{S3}           &
\colhead{Description}}
\startdata
$\mathcal{D}$	              &       $64 \times 64 \times 1024$                                 &
		  	              $128 \times 128 \times 2048$                               &
		       	              $256 \times 256 \times 4096$                               &   Domain grid size                               \\
$\mathcal{D}_A$	              &   &   $1 \times 1 \times 16$                                 &   &   Domain aspect ratio                            \\
$L$  		              &   &   $0.259$ cm $= 8 \delta_L$                              &   &   Domain width, energy-injection scale           \\
$\Delta x$                    &       $4.05\times10^{-3}$ cm                                     &
		                      $2.02\times10^{-3}$ cm                                     &
			              $1.01\times10^{-3}$ cm                                     &   Cell size                                      \\
$\widetilde{\Delta x}^{-1}$   &       $8$                                                        &
			              $16$                                                       &
			              $32$                                                       &      $\delta_L/\Delta x$                         \\
$z_{T,0}$                     &   &   $1.95$ cm $= 7.52 L$                                   &   &   Initial flame position along $z$-axis          \\
                              &   &                                                          &   &                                                  \\
$\varepsilon$                 &   &   $1.26\times10^9$ erg/cm$^3\cdot$s                      &   &   Energy-injection rate                          \\
$U_{\delta}$                  &   &   $4.53\times10^3$ cm/s $= 15 S_L$                       &   &   Turbulent velocity at scale $\delta_L$         \\
$U$                           &   &   $9.07\times10^3$ cm/s $= 30 S_L$                       &   &   Turbulent velocity at scale $L$                \\
$U_{rms}$                     &   &   $1.04\times10^4$ cm/s $= 34.48 S_L$                    &   &   Turbulent r.m.s. velocity                      \\
$U_l$                         &   &   $5.60\times10^3$ cm/s $= 18.54 S_L$                    &   &   Integral velocity                              \\
$l$                           &   &   $6.04\times10^{-2}$ cm $= 1.87 \delta_L$               &   &   Integral scale                                 \\
                              &   &                                                          &   &                                                  \\
$\tau_{ed}$                   &   &   $2.86\times10^{-5}$ s                                  &   &   Eddy turnover time, $L/U$                      \\
$t_{ign}$                     &       $3.0\tau_{ed}$                                             &
		                      $3.0\tau_{ed}$		                                 &
		       	              $2.0\tau_{ed}$		              	                 &   Time of ignition                               \\
$t_{total}$                   &   &   $16.0\tau_{ed}$                                        &   &   Total simulation time                          \\
                              &   &                                                          &   &                                                  \\
$Da$                          &   &   $0.05$                                                 &   &   Damk\"ohler number, eq.~(\ref{e:DRLG})         \\
$L_G$                         &   &   $9.47\times10^{-6}$ cm $= 2.96\times10^{-4}\delta_L$   &   &   Gibson scale, eq.~(\ref{e:DRLG})               \\
$Ma_F$                        &   &   $0.25$                                                 &   &   Mach number in fuel, $U/c_{s,F}$               \\
$Ma_P$                        &   &   $0.09$                                                 &   &   Mach number in product, $U/c_{s,P}$
\tablenotetext{a}{ Parameters common to all simulations are shown only once in S2 column.}
\enddata
\end{deluxetable}


\begin{deluxetable}{ccccc}
\tablecaption{Time-Averaged Width and Speed of the Turbulent Flame Brush\tablenotemark{a}
\label{t:Props}}
\tabletypesize{\small}
\tablewidth{0pt}
\tablehead{
\colhead{}                                    &
\colhead{$\overline{\delta_T}/\delta_L$}      &
\colhead{$O(\overline{\delta_T}/\delta_L)$}   &
\colhead{$\overline{S_T}/S_L$}                &
\colhead{$O(\overline{S_T}/S_L)$}}
\startdata
S1   &   16.13   &          &   6.09   &          \\
S2   &   14.86   &   1.96   &   4.50   &   2.29   \\
S3   &   14.42   &          &   4.09   &
\tablenotetext{a}{ Time-averaging is performed over the time
interval $(2 - 16)\tau_{ed}$.}
\enddata
\end{deluxetable}

\end{singlespace}

\clearpage

\begin{figure}[t]
\centering
\includegraphics[width=0.5\textwidth]{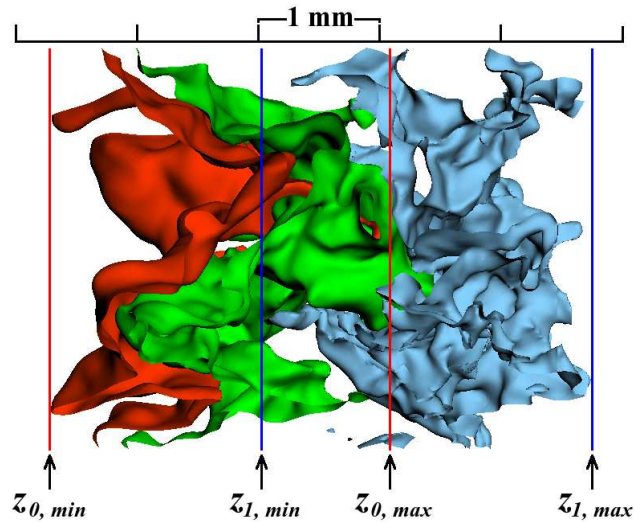}
\caption{Isosurfaces of the fuel mass fraction in simulation S2 at
$t = 13\tau_{ed}$ (cf. Fig.~\ref{f:isoV}, middle row, left
panel). Isosurface values are 0.05 (red), 0.6 (green), 0.95
(blue). Red and green isosurfaces bound the flamelet reaction zone.
Green and blue isosurfaces bound the preheat zone. The $z_{0,min}$ and
$z_{1,max}$ mark the flame-brush bounds. The $z_{0,max}$ and
$z_{1,min}$ indicate, respectively, the maximum extents of product and
fuel penetration into the flame brush.}
\label{f:isoS}
\end{figure}


\begin{figure}[t]
\includegraphics[width=1.0\textwidth]{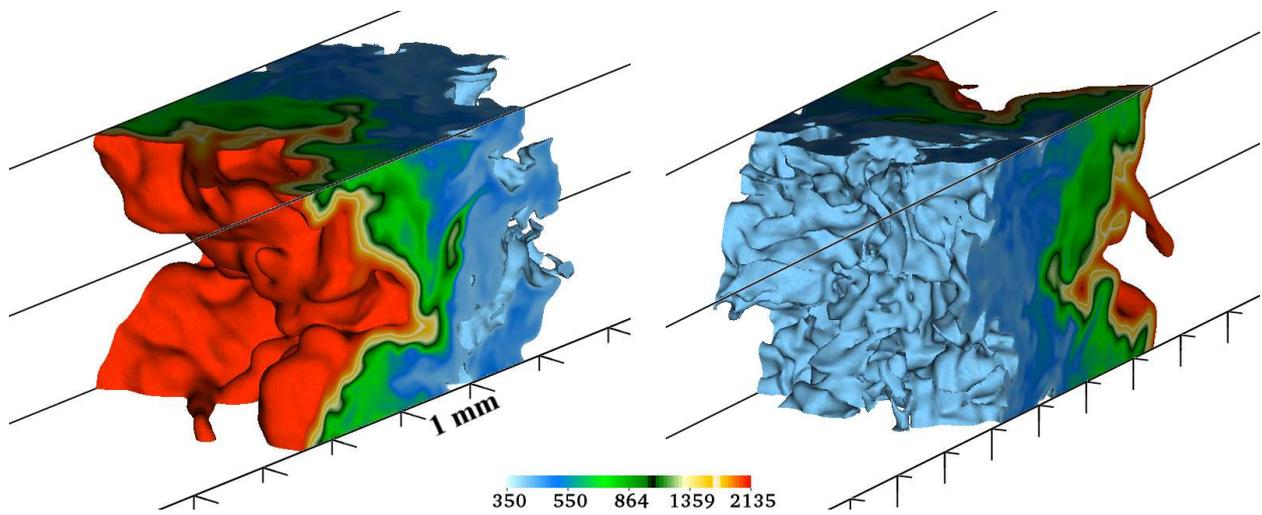}
\caption{Temperature structure of the flame brush in simulation S2
at $t = 13\tau_{ed}$ (cf. Fig.~\ref{f:isoV}, middle row). Left panel
shows view from the product side, right panel show view from the fuel
side. Bounding isosurfaces represent $T = 400$ K and $T = 2060$
K. Thin black line, corresponding to $T = 1035 K$, marks the boundary
between the preheat and reaction zones, while thin white line,
corresponding to $T = 1680 K$, shows the location of the peak reaction
rate (cf. Fig.~\ref{f:Struct}). The colormap is on a logarithmic
scale.}
\label{f:isoV.T}
\end{figure}


\begin{figure}[!t]
\centering
\includegraphics[clip, width=1.0\textwidth]{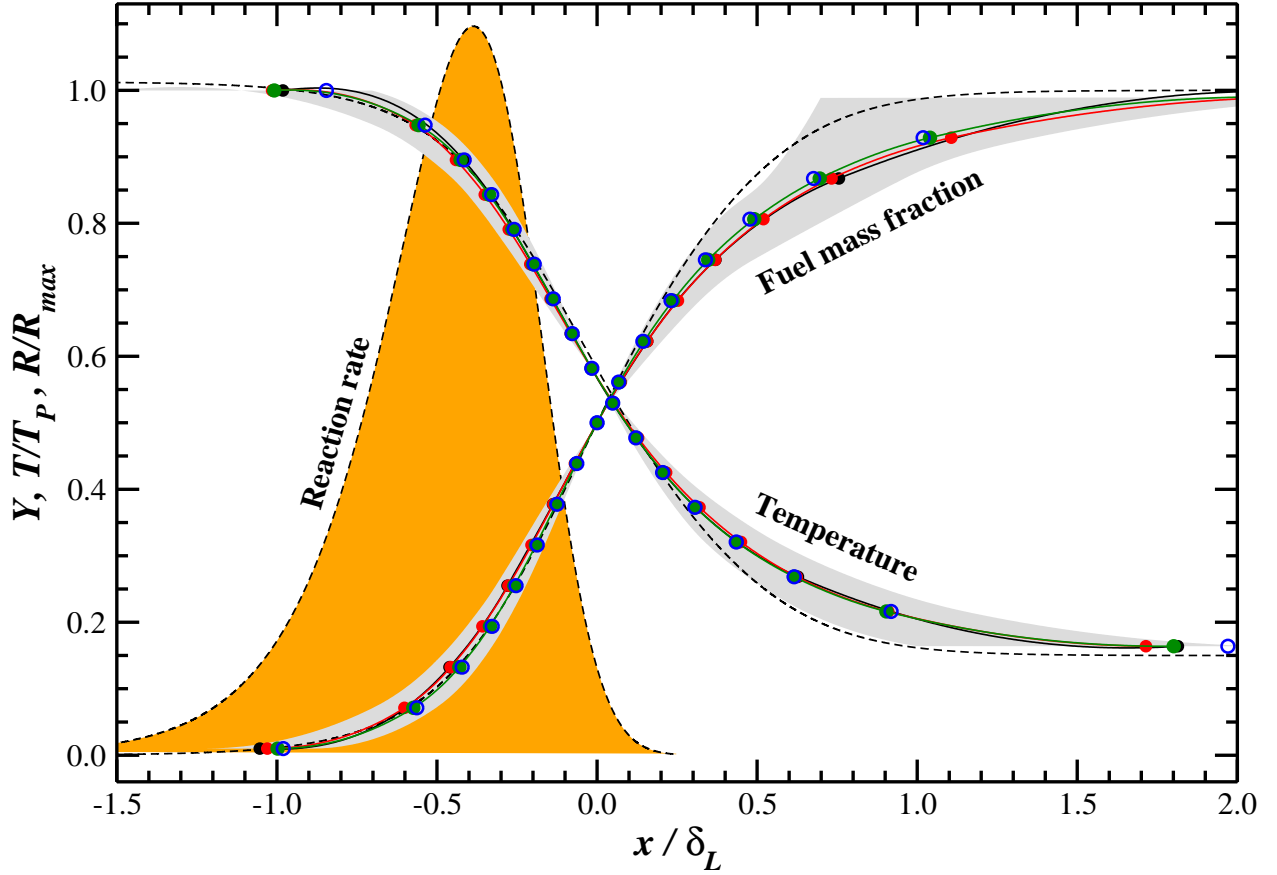}
\caption{Time-averaged flamelet structure in the turbulent flame brush.
Increasing curves represent fuel mass fraction, $Y$, decreasing curves
represent temperature, $T$. Dashed lines are exact solutions for the
planar laminar flame with the shaded orange region showing the
reaction rate, $R$. Solid black lines correspond to simulation S1, red
to S2, and green to S3. Circles represent calculated values and solid
lines are the Akima spline fits. Time averaging is performed over the
time interval $(2 - 16)\tau_{ed}$. Open blue circles represent
flamelet structure obtained in S3 using half the time averaging
interval, \ie $(9 - 16)\tau_{ed}$. Shaded gray regions show the range
of variability of individual profiles in S3 within the time-averaging
interval with all instantaneous $Y$ profiles shifted to coincide at
$Y(x = 0) = 0.5$ and $T$ profiles shifted to coincide at $(T(x=0) -
T_0)/(T_P - T_0) = 0.5$. The laminar flame structure is shown for the
fuel temperature 320 K to account for the turbulent fuel heating in
the simulation. $T$ and $R$ are normalized by their respective
reference peak values in the laminar flame with the fuel temperature
$T_0 = 293 K$, namely $T_P$ (see Table~\ref{t:Params}) and $R_{max} =
9.5 \times 10^4$ s$^{-1}$, therefore maximum values of $T$ and $R$ in
the profiles are larger than one. }
\label{f:Struct}
\end{figure}


\begin{figure}[t]
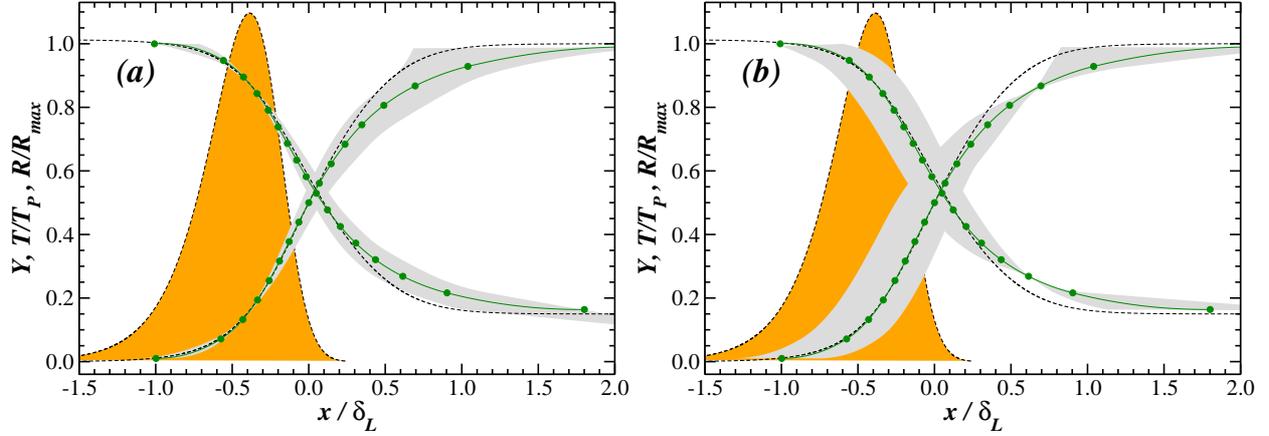

\centering 
\leavevmode
\includegraphics[clip, width=0.495\textwidth]{Flame_structure.react_zone.v15.L8.eps}
\includegraphics[clip, width=0.495\textwidth]{Flame_structure.preheat_zone.v15.L8.eps}
\caption{Variability range of instantaneous profiles of $Y$ and $T$
shifted to coincide, respectively, at $Y = 0.15$, $(T - T_0)/(T_P -
T_0) = 0.85$ (a), and $Y = 0.85$, $(T - T_0)/(T_P - T_0) = 0.15$ (b).
Same as Fig.~\ref{f:Struct} with only data for simulation S3
shown. See text for further details.}
\label{f:Variability}
\end{figure}


\begin{figure}[t]
\centering
\includegraphics[clip, width=0.5\textwidth]{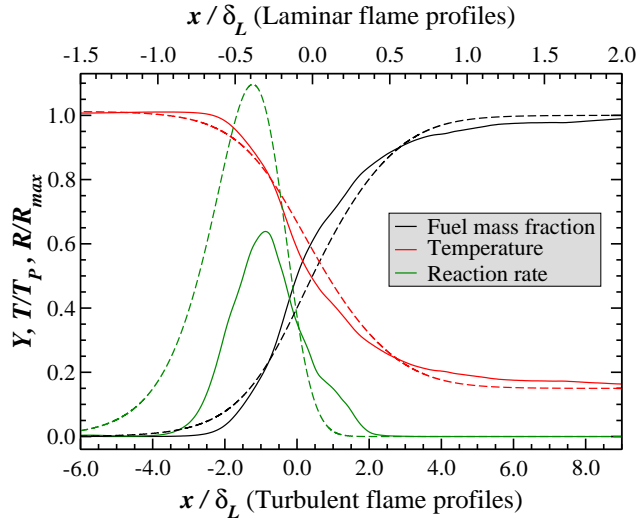}
\caption{Instantaneous average distributions of the fuel mass fraction, $Y$,
temperature, $T$, and the reaction rate, $R$, in the simulation S3 at
time $t = 12.8\tau_{ed}$. Each profile represents the distribution of
a given quantity along the $z$-axis, \ie the longest dimension of the
domain, with each point being the average over the $x-y$ plane. For
comparison, corresponding profiles for the planar laminar flame are
superimposed with dashed lines. The bottom coordinate scale is for the
turbulent flame while the top scale is for the laminar flame. Note the
difference in the range of these two scales with the turbulent flame
profiles being more than a factor of 4 wider. As in
Fig.~\ref{f:Struct}, $T$ and $R$ are normalized by $T_P$ (see
Table~\ref{t:Params}) and $R_{max} = 9.5 \times 10^4$ s$^{-1}$, the
laminar flame solution is shown for the fuel temperature 320 K, and
points of origin of the $x$-coordinates are chosen such that $Y(x = 0)
= 0.5$ and $(T(x = 0) - T_0)/(T_P - T_0) = 0.5$.}
\label{f:AverProf}
\end{figure}

\clearpage

\appendix
\section{Isosurface-based method for flamelet structure reconstruction}
\label{AppA}

Here we describe the properties of the method that we use to
reconstruct the average flamelet structure in the turbulent flame
brush, as discussed in \S~\ref{Structure}.

First consider a planar laminar flame normal to $z$-axis. In this
flame, all isosurfaces $\mathcal{S}_i$ are exactly parallel, $A_i =
Const$, and eq.~(\ref{e:isosurf}) reduces to $\eta_{i+1} \equiv
z(Y_{i+1}) - z(Y_i)$, where $z(Y)$ is based on the exact laminar flame
profile $Y(z)$. Therefore, $Y(\eta) = Y(z)$. Here for brevity we only
consider $Y$ and the same reasoning applies to $T$. Imagine now that
we deform that planar flame without stretching it, and we ensure that
at each point curvature radius is much larger than the flame width. In
this case, all isosurfaces remain exactly parallel, $A_i$ remains
constant, and again, regardless of how complex the deformation is,
this method will recover the exact laminar flame profile $Y(\eta) =
Y(z)$.

Next consider a real flame subject to the action of turbulence. In
this case, there is no reason to assume {\it a priori} either that
isosurfaces $\mathcal{S}_i$ and $\mathcal{S}_{i+1}$ are parallel or
that their surface areas are equal. Let us first define the distance
$d(p,\mathcal{S}_{i+1})$ between a point $p$, located on isosurface
$\mathcal{S}_i$, and isosurface $\mathcal{S}_{i+1}$ as
\beq
d(p,\mathcal{S}_{i+1}) = \min_{p' \in \mathcal{S}_{i+1}} \| p - p'\|_2,
\eeq
where $\|.\|_2$ is the Euclidean norm. Then we define the maximum
distance $\xi_{max}$ between the isosurfaces $\mathcal{S}_i$ and
$\mathcal{S}_{i+1}$ in a traditional way as the Hausdorff distance
\beq
\xi_{max} \equiv d_{max}(\mathcal{S}_i,\mathcal{S}_{i+1}) =
\max\Big\{\max_{p \in \mathcal{S}_i}\big(d(p,\mathcal{S}_{i+1})\big),
\max_{p' \in \mathcal{S}_{i+1}}\big(d(p',\mathcal{S}_i)\big)\Big\}.
\label{e:ximax}
\eeq
Similarly we define the minimum distance $\xi_{min}$ between
$\mathcal{S}_i$ and $\mathcal{S}_{i+1}$
\beq
\xi_{min} \equiv d_{min}(\mathcal{S}_i,\mathcal{S}_{i+1}) =
\min\Big\{\min_{p \in \mathcal{S}_i}\big(d(p,\mathcal{S}_{i+1})\big),
\min_{p' \in \mathcal{S}_{i+1}}\big(d(p',\mathcal{S}_i)\big)\Big\}.
\label{e:ximin}
\eeq
Using these definitions, the following proposition can be made:

{\it As $n \to \infty$ and $\Delta Y \equiv Y_{i+1} - Y_i \to 0$, then
$\mathcal{S}_{i+1} \to \mathcal{S}_i$ in the sense that as $\xi_{min}
\to 0$, $\xi_{max} \to \xi_{min}$.} \\
A corollary of the above proposition is the statement that $A_i
\to A_{i+1}$ as $Y_i \to Y_{i+1}$. 

In order to prove this proposition, we first make several observations.
Isosurfaces cannot intersect each other, since that would lead to
infinite gradients and unphysical conditions. Second, since the
evolution of $Y$ is governed partly by diffusion and conduction, and
there are no shocks in the system, the distribution of $Y$ is smooth
and continuous. Moreover, $Y$ cannot be exactly constant over any
extended region since advective and diffusive processes in an unsteady
turbulent flow would inevitably create spatial variations in the
distribution of $Y$ on all scales. Therefore, isosurfaces
corresponding to all values $Y_i < Y < Y_{i+1}$ are all bounded by the
isosurfaces $\mathcal{S}_i$ and $\mathcal{S}_{i+1}$, they are smoothly
distributed throughout the volume bounded by $\mathcal{S}_i$ and
$\mathcal{S}_{i+1}$, and as $Y_i
\to Y_{i+1}$, $\xi_{min} \to 0$.

Let us first prove that a value $Y'\in (Y_i,Y_{i+1})$ can always be
found such that for its isosurface $\mathcal{S}'$ the following is
true: $\xi_{min} - \xi_{min}' \leq \xi_{max} - \xi_{max}'$. Here
$\xi_{max}' \equiv d_{max}(\mathcal{S}',\mathcal{S}_{i+1})$ and
$\xi_{min}' \equiv d_{min}(\mathcal{S}',\mathcal{S}_{i+1})$ by analogy
with eqs.~(\ref{e:ximax})-(\ref{e:ximin}). Assume the contrary, namely
that for all values of $Y'$ from the interval $(Y_i,Y_{i+1})$ we have
$\xi_{min} - \xi_{min}' > \xi_{max} - \xi_{max}'$. This means that as
$Y' \to Y_{i+1}$, $\xi_{min}' \to 0$ and $\xi_{max}' >
\xi_{max} - \xi_{min}$. Therefore, $\mathcal{S}'$ isosurfaces have
some limiting isosurface $\mathcal{S}^*$ which has the same value of
$Y^* = Y_{i+1}$ but which is distinct from $\mathcal{S}_{i+1}$ because
$\xi_{max}^* > \xi_{max} - \xi_{min}$. Since isosurfaces do not
intersect, this means that no other isosurfaces pass through the
volume bounded by $\mathcal{S}^*$ and $\mathcal{S}_{i+1}$.
Consequently, $\mathcal{S}^*$ and $\mathcal{S}_{i+1}$ bound a region
of constant $Y$. We discussed above, however, that such regions cannot
exist. Therefore, we arrive at a contradiction and our assumption was
wrong, thus proving our initial statement.

It then follows, that there exists a value $Y'' \in (Y',Y_{i+1})$ such
that for its isosurface $\mathcal{S}''$ we have $\xi_{min}' -
\xi_{min}'' \leq \xi_{max}' - \xi_{max}''$. By induction this proves
the original proposition. A similar result can be proven for
temperature.

In practice, this result means the following. The value of $\xi_{max}
- \xi_{min}$ is the measure of the variation of $\mathcal{S}_i$ with
respect to $\mathcal{S}_{i+1}$. As we select finer discretization
intervals of $Y$, this variation decreases while $\xi_{min} \to 0$.
Therefore, the isosurfaces become more and more parallel to each
other. An illustration of this can be seen in Fig.~\ref{f:isoS}. While
there is very little in common between the $Y = 0.05$ and $Y = 0.95$
isosurfaces, $Y = 0.6$ tends to follow the $Y = 0.05$ isosurface much
more closely.

An important question concerns the choice of the number of
discretization intervals $n$ in practical applications. It follows
from the above discussion, that in order to maximize the accuracy of
the method, $n$ must be chosen as large as possible. This ensures that
consecutive isosurfaces are close to each other and that $\xi_{max} -
\xi_{min}$ is minimal or, ideally, zero. In the computational domain,
the minimum spatial scale is set by the cell size $\Delta x$. This
determines the minimum practical separation of isosurfaces
$\xi_{min}$. Isosurfaces with smaller separations would pass through
the same cell, which would result in a substantial drop in accuracy of
the overall method since all flow variables are piecewise constant
within a cell. At the same time, if two isosurfaces on average have
separation of a few $\Delta x$, this ensures that their surface areas
are close in value, and there cannot exist an intermediate isosurface
which is substantially more or less tightly folded. These
considerations allow us to determine the maximum value of $n$.

If we assume that the maximum gradient in temperature in the domain is
close to that in the laminar flame profile $(dT/dx)_{L,max}$, then
\beq
\frac{\Delta T_{min}}{(dT/dx)_{L,max}} = \xi_{min} \approx \Delta x.
\label{e:dT}
\eeq
Recalling that $\delta_L = (T_P - T_0)/(dT/dx)_{L,max}$, we can
rewrite eq.~(\ref{e:dT}) as
\beq
\frac{T_{max} - T_{min}}{n} = \frac{T_P - T_0}{\delta_L}\Delta x,
\eeq
where $T_{min}$ and $T_{max}$ are the lower and upper bounds of the
discretized temperature range. Since $T_{min} \approx T_0$ and
$T_{max} = T_P$, we find that
\beq
n_{max} \sim \frac{\delta_L}{\Delta x}.
\label{e:n}
\eeq
A similar argument applies to the fuel mass fraction. Therefore, $n$
should not be larger than the number of grid cells within the laminar
flame thermal width. Typically, we find the choice of $n$ based on
eq.~(\ref{e:n}) adequate in practical applications.

We found that adaptivity of the algorithm, which implements the method
described here, is essential to guarantee high accuracy. The estimate
given by eq.~(\ref{e:n}) assumed that the largest gradients present in
the system are well approximated by $(dY/dx)_{L,max}$ and
$(dT/dx)_{L,max}$ in the laminar flame profile. This may not be the
case. Therefore, it is important for the algorithm to be able to
decrease $n$ if larger gradients are encountered. On the other hand,
the choice of $n$ in eq.~(\ref{e:n}) is based on the thermal flame
width. The full flame width is larger than $\delta_L$. Therefore,
while eq.~(\ref{e:n}) is adequate to capture flame structure in the
steepest region, the parts of the profile close to the extreme values
of the discretized range may be under-resolved (see
Fig.~\ref{f:Struct}). In those regions, both $Y$ and $T$ vary slowly in
space. Thus, for consecutive isosurfaces, $\eta_i \gg \Delta x$.
Ideally, $\Delta Y$ and $\Delta T$ must be chosen such that $\Delta x
\lesssim \eta_i \lesssim \alpha \Delta x$ for $i \in [1,n]$. Typically,
we find that $\alpha = 4$ is a reasonable choice.

Then the adaptivity of the algorithm is implemented as follows.  The
value of $n$ is adjusted to ensure that both $\Delta Y$ and $\Delta T$
are marginally large enough to prevent situations when $\eta_i <
\Delta x$. On the other hand, if $\eta_i > \alpha \Delta x$ is
encountered, $\Delta Y$ or $\Delta T$ for that interval is divided by
two. Subsequently, $\eta$ is evaluated separately for the first and
second half-intervals, and the final $\eta_i$ is the sum of those
values. If any of the intermediate values of $\eta$ is larger than
$\alpha \Delta x$, that half-interval is again subdivided by two and
the whole procedure proceeds recursively until on each of the substeps
the condition $\eta < \alpha \Delta x$ is satisfied. This ensures that
$\eta$ is always determined for isosurfaces which are close to each
other.

Note, that this algorithm uses the laminar flame structure only to
provide the initial guess for $n$. Subsequently, it adapts the
discretization intervals for $Y$ and $T$ based only on the actual
gradients found in the flow. Therefore, information about the laminar
flame structure is not required at all, and eq.~(\ref{e:n}) only
facilitates finding the correct $\Delta Y$ and $\Delta T$. At the same
time, in situations when the internal structure of the turbulent flame
is very different from that of the laminar flame, as would be expected
in the case of distributed burning, eq.~(\ref{e:n}) is only of limited
benefit.

A key limitation of this method is that it can be applied only to
quantities that change monotonically through the flame, such as $Y$,
$T$, or $\rho$. In particular, this procedure cannot be used to
directly determine the distribution of the reaction rate, $R$.
Non-monotonic behavior of a quantity means that there is an inherent
degeneracy in its distribution inside the flame. Consider a planar
laminar flame (\eg see Fig.~\ref{f:Struct}). Two distinct points in
the reaction rate profile have the same value of $R$.  Therefore,
there would be two isosurfaces for each value of $R_i$ and $R_{i+1}$.
Volume $V_i$ would be the total volume bounded by both pairs of
isosurfaces and area $A_i$ would be the combined surface area of both
isosurfaces $\mathcal{S}'_i$ and $\mathcal{S}''_i$. There is no
mechanism in this method to distinguish the contributions of each pair
of isosurfaces both into $A_i$ and $V_i$. Consequently, $\eta_i$ found
using eq.~(\ref{e:isosurf}) would have no meaning.

In general, we find that the method described above performs best for
quantities that predominantly change within the volume of the flame
brush, and that do not vary at all, or vary only slightly around the
limiting values of the discretized range, outside the flame brush.
This is the case for $Y$ and $T$. The accuracy is the highest for $Y$
which is exactly 0 or 1 outside the flame brush. Therefore, results
are not contaminated by the isosurfaces which pass outside the burning
region. For temperature, the accuracy is marginally lower due to the
fluctuations in the turbulent field away from the flame brush, which
can cause values of $T$ to become slightly below the upper bound or
above the lower bound. As a result, isosurfaces would pass through
such regions which are not part of the flame brush. Typically,
however, this can introduce small errors only near the extreme values
in the temperature profile. At the same time, for a quantity like
density, the accuracy drops to the point that the obtained profiles
cannot be reliably used to analyze the density structure of the
flamelets. This is primarily due to the fact that variations in
density outside the flame brush are large enough to be significant in
comparison with the change in $\rho$ inside the flame brush
itself. Therefore, a large portion of the profile ends up being
contaminated by the contributions from the regions which are outside
the flame brush, but which happen to have values of $\rho$ that fall
within its discretized range.

Finally, this method does not provide any means to relate the
distributions of different quantities, \eg $Y$ and $T$. In a planar
laminar-flame structure, for instance, a given value of $Y$ uniquely
defines its position $z$ in the profile, and thus it uniquely
determines the value of $T = T(z)$. This cannot be done for the
profiles obtained using the method described above. Recall that $\eta$
is not a spatial coordinate but simply the distance between
consecutive isosurface values, and it is assumed that $\eta_0 = 0$.
Thus, values of $\eta$ for $Y$ and $T$ are distinct and unrelated and
other arguments must be invoked in order to relate their
distributions.


\end{document}